\documentclass{article}
\usepackage[utf8]{inputenc}
\usepackage{natbib}
\usepackage{graphicx}
\graphicspath{ {./images/} }
\usepackage{nicefrac}       
\usepackage{algorithm} 
\usepackage{algorithmicx} 
\usepackage{algpseudocode}
\usepackage{bm}
\usepackage{amsmath, amsthm}
\usepackage{amsfonts}
\usepackage{multirow}
\usepackage{url}
\usepackage{array}
\usepackage{hyperref}
\usepackage{booktabs}
\usepackage[letterpaper, margin=1in]{geometry}

\usepackage{authblk}
\usepackage{float}

\renewcommand{\vec}[1]{\bm{#1}}

\DeclareMathOperator*{\argmax}{arg\,max}
 
\newcommand*{\bfrac}[2]{\genfrac{}{}{0pt}{}{#1}{#2}}
\newcommand{\botrule}{\toprule}
\newcommand{\titleStr}{Beyond similarity assessment: \\ Selecting the optimal model for sequence alignment via the Factorized Asymptotic Bayesian algorithm}

\title{\titleStr}
\author[1]{Taikai Takeda\thanks{The author is now at Indeed, Recruit Holdings Co., Ltd.}}
\author[1,2,3,4,5]{Michiaki Hamada\thanks{Corresponding author: \href{mailto:hamada@waseda.jp}{mhamada@waseda.jp}}}

\affil[1]{Department of Electrical Engineering and Bioscience, Waseda University}
\affil[2]{Computational Bio Big-Data Open Innovation Laboratory (CBBD-OIL), National Institute of Advanced Industrial Science and Technology (AIST)}
\affil[3]{Artificial Intelligence Research Center (AIRC), National Institute of Advanced Industrial Science and Technology (AIST)}
\affil[4]{Institute for Medical-oriented Structural Biology, Waseda University}
\affil[5]{Graduate School of Medicine, Nippon Medical School}
\date{}

\begin{document}

\maketitle

\begin{abstract}
Pair Hidden Markov Models (PHMMs) are probabilistic models used for pairwise sequence alignment, a quintessential problem in bioinformatics. PHMMs include three types of hidden states: match, insertion and deletion. Most previous studies have used one or two hidden states for each PHMM state type. However, few studies have examined the number of states suitable for representing sequence data or improving alignment accuracy. We developed a novel method to select superior models (including the number of hidden states) for PHMM. Our method selects models with the highest posterior probability using Factorized Information Criteria (FIC), which is widely utilised in model selection for probabilistic models with hidden variables. Our simulations indicated this method has excellent model selection capabilities with slightly improved alignment accuracy. We applied our method to DNA datasets from 5 and 28 species, ultimately selecting more complex models than those used in previous studies.
\end{abstract}

\section{Introduction}
The alignment of biological sequences (e.g. DNA, RNA and proteins) is one of the most classical and important problems in the field of bioinformatics. Sequence alignment permits the assessment of the functional relationships among biological sequences by quantifying sequence similarity. 
Because similar nucleotides or amino acids sequences are often functionally related, the development of quantitative evaluations of sequence similarity has been of great interest. This high demand for similarity evaluations has driven the development of a variety of alignment programs \citep{Altschul1990,Thompson1994,Frith2010}.
Moreover, sequence alignments are essential for analysing the huge amounts of sequence data produced by high-throughput sequencers in computational tasks such as mapping read sequences onto reference genomes \citep{pmid20460430,pmid28039163}.

For this alignment task, probabilistic approaches are widely recognised. These probabilistic approaches include Pair Hidden Markov Models (PHMMs) \citep{Durbin1998}, which handle indels and substitutions that occur throughout molecular evolution by using sequentially dependent unobserved hidden states, specifically the match, insertion and deletion states as well as their corresponding probabilistic symbol emissions (cf. Figure~\ref{fig:hidden_states}). 

There have been several attempts to construct slightly more complex PHMMs.

\citet{Lunter2008,pmid19478997,pmid18849524} used PHMMs with two insertion and two deletion states, and \citet{pmid19042944} proposed a general version of PHMMs that employs a zeta power-law model of indel lengths.
Additionally, a few generalisations of PHMMs have been proposed, such as \citet{pmid12015888}, which introduced generalised PHMMs for DNA--DNA, DNA--cDNA and DNA--protein alignments.
However, to the best of our knowledge, no previous study has focused on determining the suitable number of states for representing the biological models that describe sequence evolution or for achieving better alignment accuracy.


Bayesian model selection provides a sophisticated approach for selecting the best model by maximising model evidence. In this, some parameters are marginalised out, and so a preference for simpler models is inherent to the method. When the model prior is uniform, maximising model evidence is equivalent to maximising the posterior probability of model given data, so we can choose the model with the largest posterior by maximising the model evidence.


The well-known difficulty of Bayesian model selection is that the model evidence is analytically intractable in general,
including for PHMMs.
Markov Chain Monte Carlo (MCMC) \citep{Hastings1970} and Variational Inference (VI) \citep{Jordan1999,Beal2003,Blei2016} enable approximation of the difficult-to-compute model evidence, but both approaches have a drawback: 
high computational cost.
In contrast, the Factorized Asymptotic Bayesian (FAB) algorithm \citep{Fujimaki2012,Fujimaki2012a,Hayashi2015} is a promising alternative model selection technique based on the Factorized Information Criterion (FIC). 
One advantage of the FAB algorithm is its simultaneous optimisation of the model structure and the parameters, which makes the FAB algorithm more scalable than VI and MCMC. The advantages are further  discussed in section \ref{sec:FAB-PHMM}.

The contributions of this study are summarised as follows.
\begin{enumerate}
\item We developed a novel FIC-based model selection algorithm for PHMMs and demonstrate the reasonably good accuracy in model selection using a synthetic dataset. To the best of our knowledge, this is the first attempt in the literature to apply a model selection method to PHMMs.
\item The model selection method slightly improved evaluation metrics on the same synthetic dataset.
\item We conducted experiments on real DNA sequences and found that our method selects a more complex probabilistic structure than the ones that have been traditionally used for pairwise alignment of these species. 
\end{enumerate}


\section{Methods}
\begin{figure}[tb]
\centering
\centerline{\includegraphics[width=.6\columnwidth]{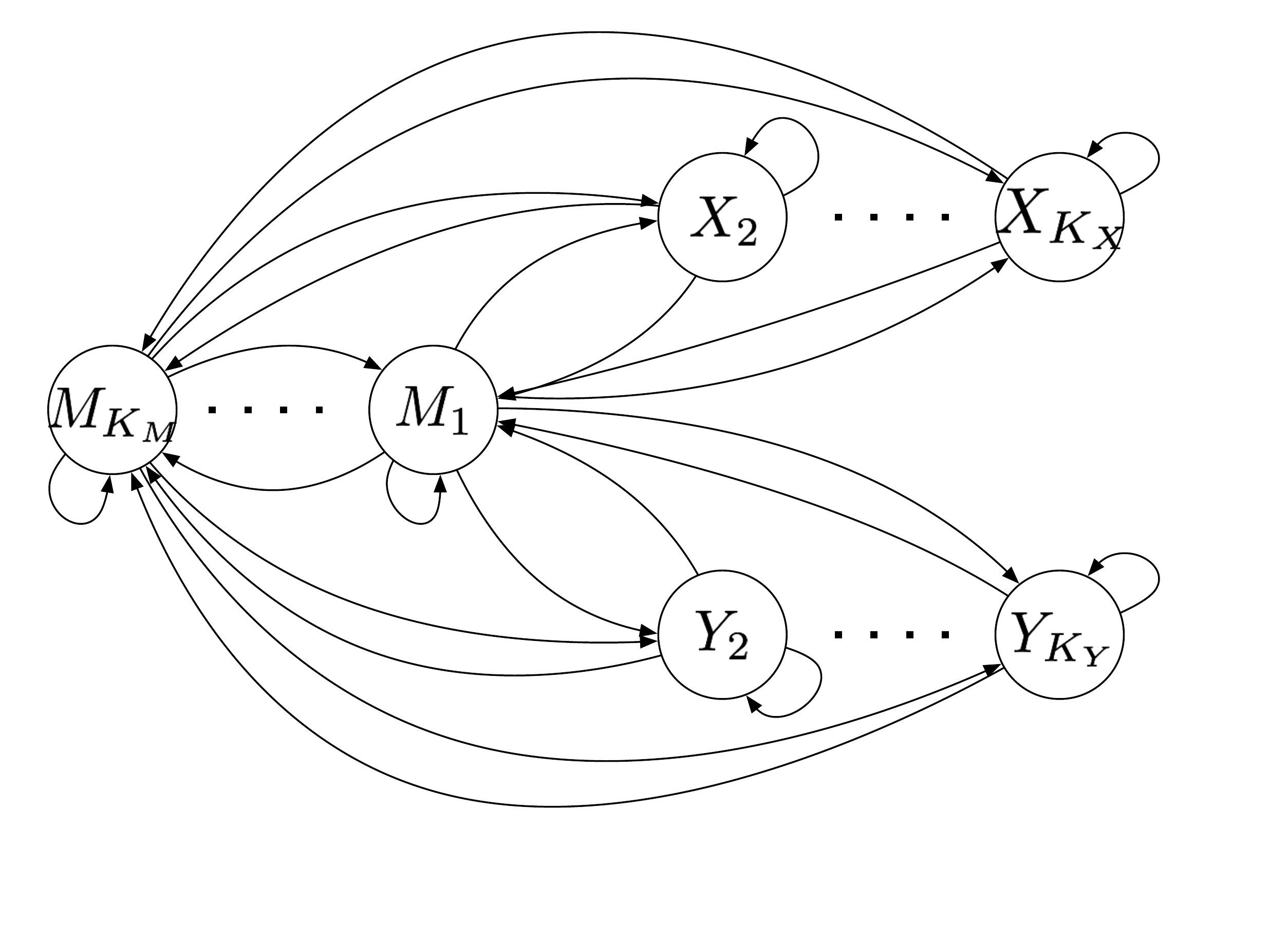}}

  \caption[Transition diagram of PHMM]{Transition diagram of hidden states in a Pair Hidden Markov Model (PHMM). The match states, which emit a pair of characters ($M_1\ldots M_{K_M}$), are connected to all the other states whereas the X- and Y-insertion states, which emit a pair of a character and a gap symbol `-' ($X_1\ldots X_{K_X}$ and $Y_1\ldots Y_{K_Y}$), are only connected to the Match states.}{}
\label{fig:hidden_states}
\end{figure}

PHMMs are a type of probabilistic generative model for sequence alignment \citep{Durbin1998}
with three types of hidden states: a match-type state $M$, an X-insertion-type state $X$ and a Y-insertion-type state $Y$ (Figure \ref{fig:hidden_states}). The insertion states model the molecular evolution of indels, and the emission probability of the match states characterises the substitution rates.

In this study, we employed the FAB algorithm \citep{Fujimaki2012, Hayashi2015} {to select the best model structure for a PHMM}. The FAB algorithm is an information criterion-based technique that enables the simultaneous optimisation of both the parameters and the model structure. {The properties of the FAB algorithm are explained in \ref{sec:FAB-PHMM} in more detail.}
Note that we modified the standard formalisation of PHMM (e.g. \citet{Durbin1998}) because it does not use hidden variables explicitly, which is inappropriate for the FAB algorithm. In this section, we introduce our formalisation of the PHMM with explicit hidden variables (Section~\ref{sec:PHMM}) and then develop a proposed model selection method using the FAB algorithm (Section~\ref{sec:FAB-PHMM}).

In the following, we denote the number of match-type states, X-insertion-type states and Y-insertion-type states as $K_M$, $K_X$ and $K_Y$, respectively. Additionally, $K$ represents the total number of hidden states, that is $K=K_M+K_X+K_Y$. Formally, we regard model selection as selecting the number of hidden states $(K_M, K_X, K_Y)$. 

\begin{figure}[tb]
\centerline{
\begin{tabular}{cc}
(a) Hidden Markov Model (HMM)  & (b) Pair Hidden Markov Model (PHMM)\\
\includegraphics[width=0.5\linewidth]{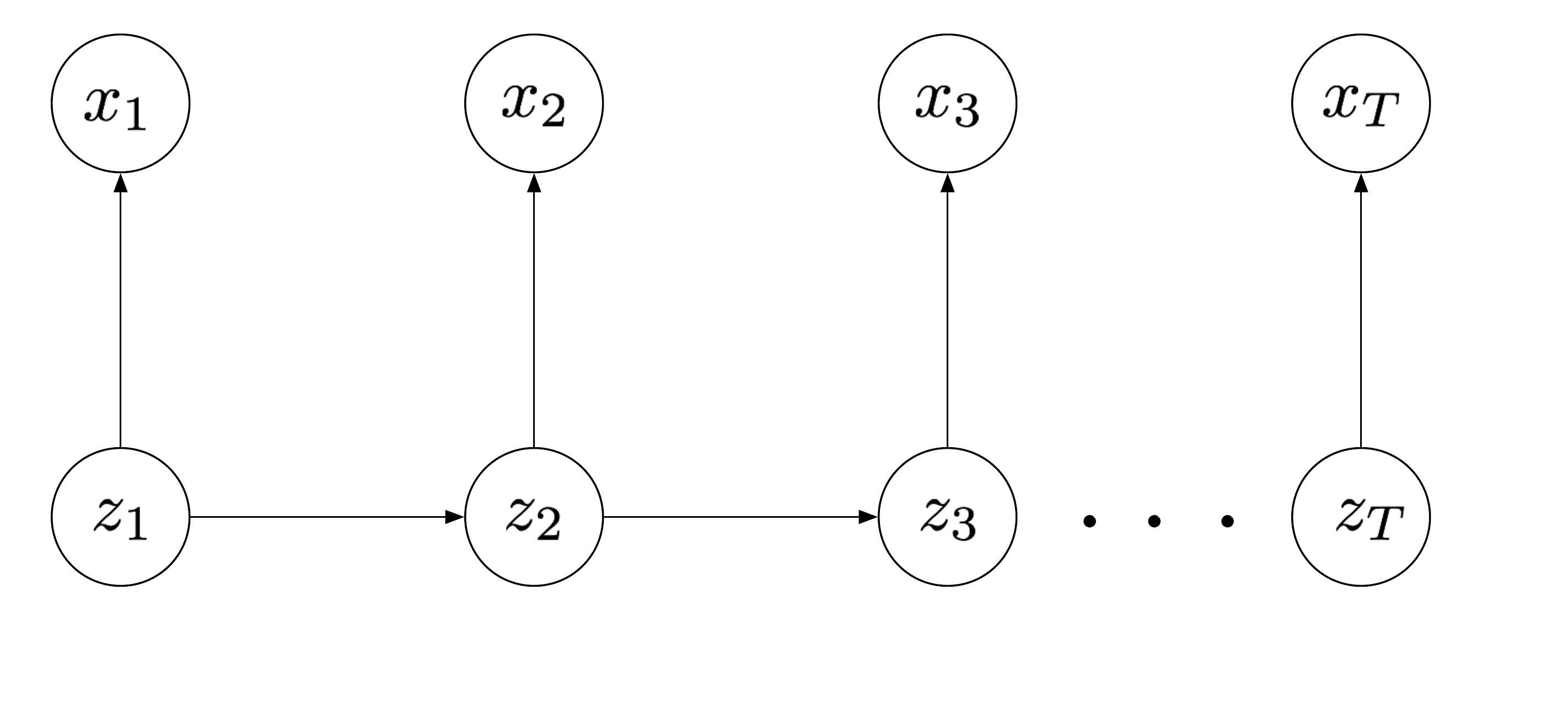} & 
\includegraphics[width=0.5\linewidth]{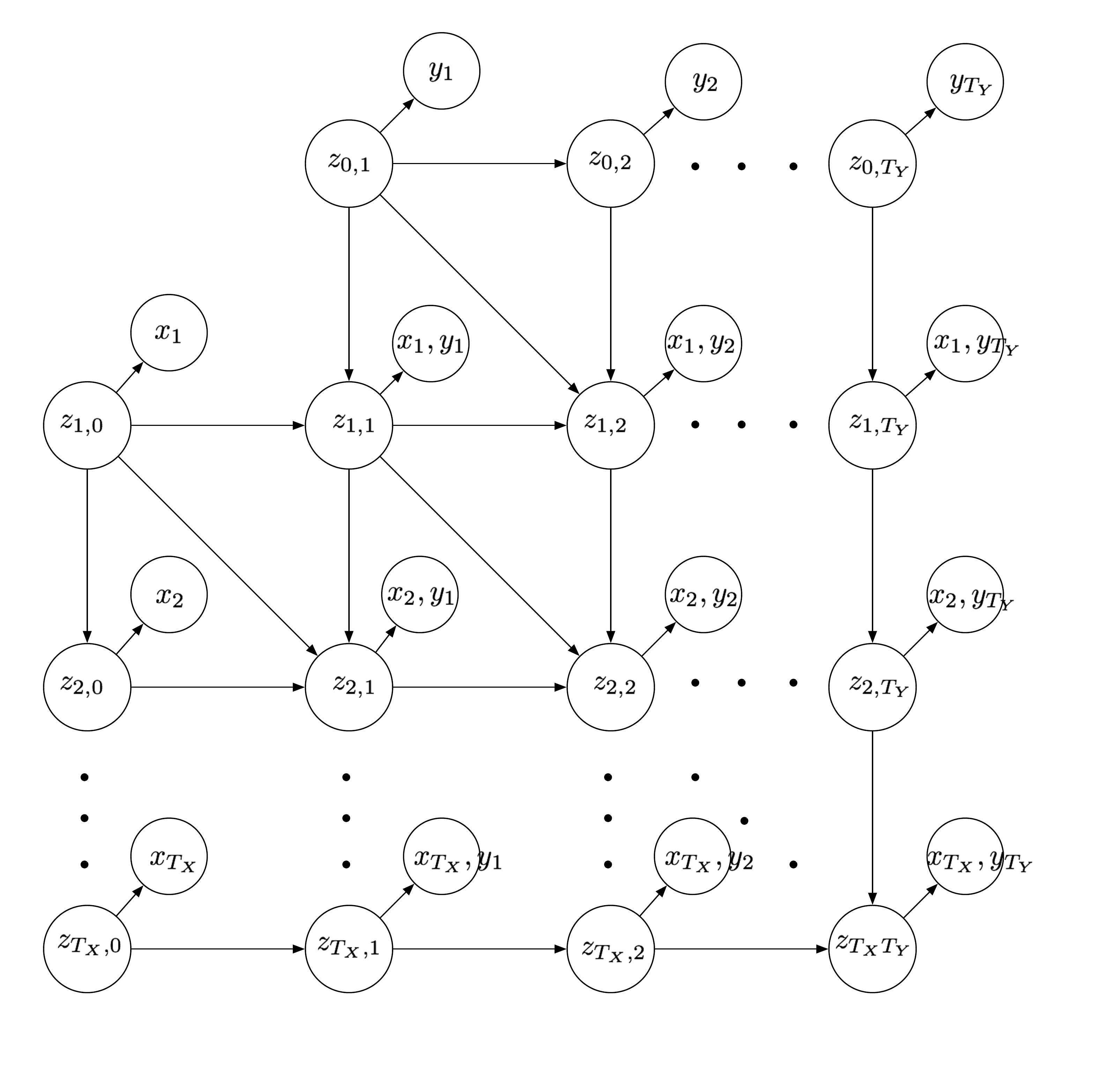}
\end{tabular}}
\caption[Graphical models of HMM and PHMM]{Graphical model representation of (a) HMMs and (b) PHMMs. The hidden states, denoted by $z_t$, are one-dimensional in the normal HMM, whereas they, denoted by $(z_{t,u})$, are two-dimensional in the PHMM.
In the PHMM, a pair of symbol emissions $(x_t, y_u)$ is an emission from the hidden state $z_{t,u}$, describing a pair of (aligned) nucleotides in the case of DNA alignments, for example.}
\label{fig:gm_hstates}
\end{figure}

\subsection{Pairwise Hidden Markov Model (PHMM)}\label{sec:PHMM}
\begin{figure}[tb]
\centerline{
\includegraphics[width=0.6\linewidth]{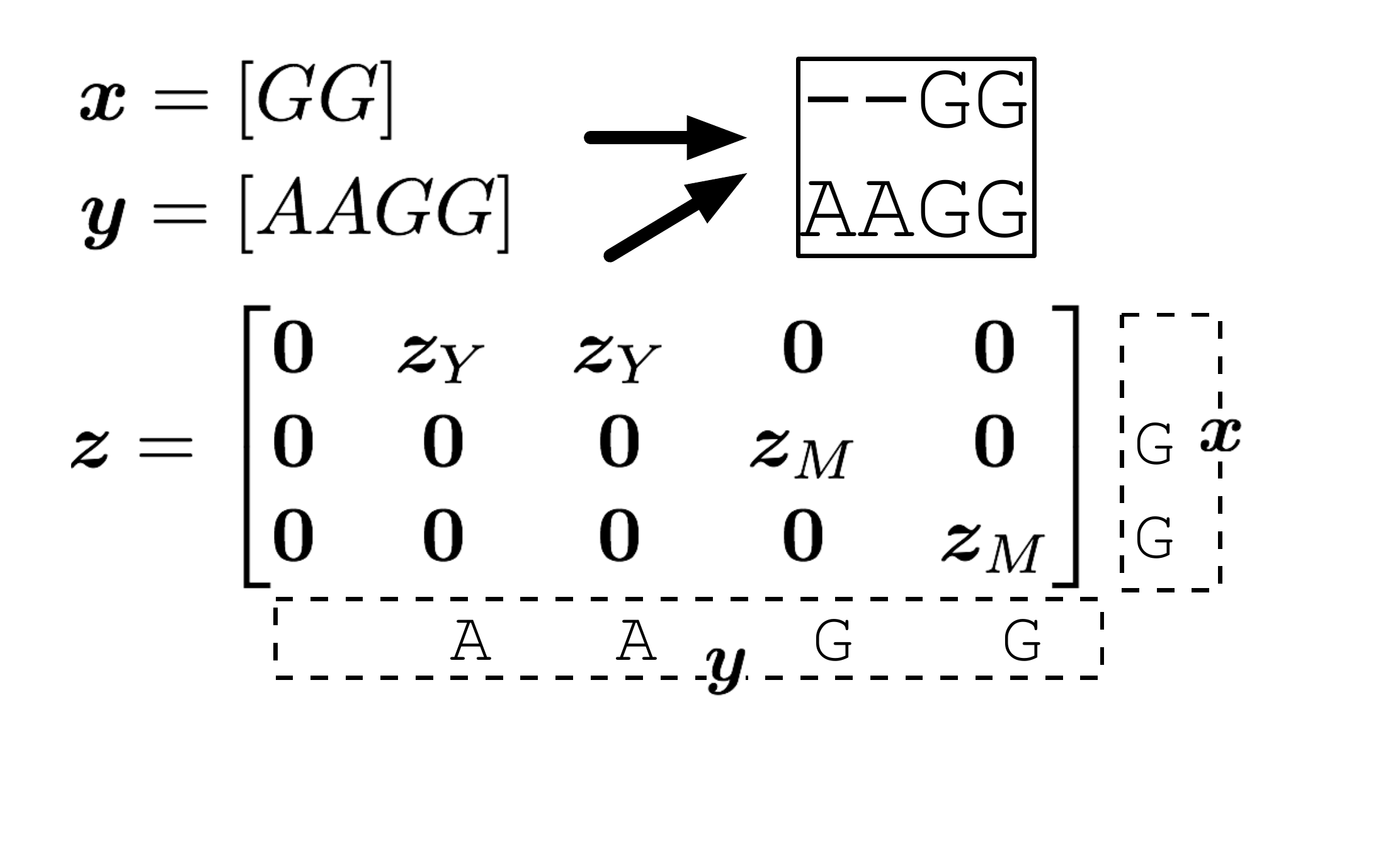}
}
\caption{An alignment example for a pair of DNA sequences $x$ and $y$ as well as the corresponding \textit{two-dimensional} hidden states, where $(K_M, K_X, K_Y)=(1,1,1)$, illustrating how hidden states are encoded. $\vec{0} = (0,0,0)$ is the zero-state that does not emit any symbols, and $\vec{z}_M = (1, 0, 0)$, $\vec{z}_X = (0,1,0)$ and $\vec{z}_Y = (0,0,1)$ are the $1$-of-$K$ coding corresponding to the $M$-type, $X$-type and $Y$1-type states, respectively. This hidden state encoding allows us to generate the alignment $\bfrac{--GG}{AAGG}$ from the sequence pair $GG$ and $AAGG$.}
\label{fig:hstates_example}
\end{figure}

Let observed sequences be $\vec{x} = \{\vec{x}^n\}_{n \in [1,N]}$ and $\vec{y} = \{\vec{y}^n\}_{n \in [1,N]}$, where $N$ is the number of sequence pairs. 
The $n$-th sequences are $\vec{x}^n = \{x^n_t\}_{t\in [1,T^n_X]}$ and $\vec{y}^n = \{y^n_u\}_{u\in [1,T^n_Y]}$, where $T_{X}^n$ and $T_{Y}^n$ are the lengths of $\vec{x}^n$ and $\vec{y}^n$, respectively. We abbreviate all the observed sequences as $\vec{X} = \{ \vec{x}, \vec{y} \}$.
Unlike normal HMMs, PHMMs have hidden variables $\vec{Z} = \{\vec{z}^n\}_{n\in[1, N]}$, which are \textit{two-dimensional} and the $n$-th of which is $\vec{z}^n = \{\vec{z}^n_{tu}\}_{t \in [0, T^n_X], u \in [0, T^n_Y]}$ (Figure \ref{fig:gm_hstates}). Note that these two-dimensional hidden variables are not a common formalisation and include a \textit{zero-state}, introduced below. The value $\vec{z}^n_{t, u}$ corresponds to the hidden state where, for match states, $x^n_t$ and $y^n_u$ are matched.
For insertion states, $\vec{z}^n_{tu}$ represents that $x_t$ corresponds to the gap ``--'' and that the last-used symbol in $\vec{y}^n$ is $y^n_u$ and vice versa for Y-insertion states. The hidden state $\vec{z}_{tu}^n = \{ z_{t,u, k}^n\}_{k \in [1,K]}$ is a 1-of-$K$ representation, but slightly modified to allow a \textit{zero-state}, where $z^n_{t,u, k}$ for all $k$ is zero and does not emit any symbols from that variable (an example is shown in Figure \ref{fig:hstates_example}). 
This is because of the unique characteristics of PHMMs (in comparison to conventional HMMs); only a subset of the hidden variables emit symbols, that is only the hidden variables corresponding to aligned positions emit symbols (Figure \ref{fig:hstates_example}).
The set $\vec{\Pi} = \{\vec{\alpha}, \vec{\beta}, \vec{\phi}\}$ is a parameter set, where $\vec{\alpha}, \vec{\beta}$ and $\vec{\phi}$ represent the initial probability, transition probability, and emission probability parameters, respectively. 
 Also, each hidden state $k$ corresponds to one of the state types $\{M, X, Y\}$, which is given by a function $\mathcal{S}$ where $\mathcal{S}(k) \in \{M, X, Y\}$.

Now we can write the complete log-likelihood of PHMM as
\begin{align}
\ln  p (\vec{X}, \vec{Z}|\vec{\Pi}) =\sum_{n=1}^N \bigg[ \ln p(\vec{z}^n_{\text{in}}| \vec{\alpha}) + 
\sum_{t=0}^{T_X^n} \sum_{u=0}^{T_Y^n} \Big( \ln p(\vec{z}^n_{tu} | pa(\vec{z}^n_{tu}), \vec{\beta}) + \ln p(x^n_t, y^n_u | \vec{z}^n_{t u}, \vec{\phi}) \Big) \bigg] \label{eq:phmm_ll_short}
\end{align}
where $\vec{z}^n_{\text{in}}$ is a set of hidden variables corresponding to the initial states. The initial hidden variable varies with the type of hidden state because each hidden state's type uses a different number of original sequences; an $M$-type state uses both $x_1$ and $y_1$, while $X$- and $Y$-type states each use one of them. For this reason, the initial hidden variable of an $M$-type state is $\vec{z}_{1,1}$, whereas it is $\vec{z}_{1,0}$ and $\vec{z}_{0, 1}$ for $X$-type and $Y$-type states, respectively. For the transition probability, we use $pa(\vec{z}^n_{tu}) = \{ z_{t^\prime u^\prime k} \}_{k \in [1,K]}$ as a set of (previous) hidden variables from which this can transit to $\vec{z}_{t, u}$. 
We further denote the emission probability as a categorical distribution using new variables $\{ \psi^n_{tuk} \}$ as
\begin{eqnarray}
p(x_t, y_u | z^n_{tuk} = 1) = \psi^n_{tuk} = 
\begin{cases}
\phi_k (x_t, y_u) \quad \text{if} \quad \mathcal{S}(k) = M \\
\phi_k (x_t, -) \quad \text{ if} \quad \mathcal{S}(k) = X \\
\phi_k (-, y_u) \quad \text{ if} \quad \mathcal{S}(k) = Y
\end{cases}
\end{eqnarray}
where $\phi_k$ represents the categorical emission probability of the $k$-th hidden state. Note that the $X$- and $Y$-type states emit the gap ``--'' instead of the normal symbols (e.g. A, T, G or C in the case of DNA alignments). Thus, the dimensionality of the parameter of the emission probability differs with the type of hidden states, namely, $L^2 - 1$ for the match states and $L-1$ for the insertion states, where $L$ is the number of symbols ($L=4$ in the case of DNA sequences). Using this notation, we can rewrite the complete likelihood in an explicit form.
{
\begin{align}
\ln p (\vec{X}, \vec{Z}|\vec{\Pi}) 
&=
\sum_{n=1}^N \bigg[ \ln p(\vec{z}^n_{\text{in}}| \vec{\alpha})+ 
\sum_{t=0}^{T_X} \sum_{u=0}^{T_Y} \sum_{k=1}^K \Big( 
{pa(\vec{z}^n_{tu})_k} \ln p_k(\vec{z}^n_{tu} | \vec{\beta}_k)
+ {z^n_{tuk}} \ln p(x^n_t, y^n_u | \vec{z}^n_{t u}, \vec{\phi}_k)
\Big) \bigg] \nonumber\\
&=
\sum_{n=1}^N \sum_{k=1}^K \bigg[
z^n_{d_{xk}, d_{yk}, k}\ln \alpha_k +
\sum_{t=0}^{T_X^n} \sum_{u=0}^{T_Y^n} \Big(
 \sum_{l=1}^K 
z^n_{(t - d_{xk}),(u - d_{yk}), k}z^n_{t u l} \ln \beta_{kl} 
+ z^n_{tuk} \ln \psi_{tuk}^n  \Big) \bigg]\label{eq:likelihood}
\end{align}
where $p_k(\vec{z}_{tu} | \vec{\beta}_k) = \prod_{l=1}^K p(z_{tul} = 1 | z_{(t - d_{xl}), (u - d_{yl}), k} = 1)^{z_{tul}}$
and $(d_{xk}, d_{yk})$ is a transition direction defined as
}
\begin{align*}
(d_{xk}, d_{yk}) = 
\begin{cases}
(1,1) \quad \text{if}\quad \mathcal{S}(k) = M\\
(1,0)  \quad \text{if}\quad \mathcal{S}(k) = X\\
(0,1)  \quad \text{if}\quad \mathcal{S}(k) = Y.
\end{cases}
\end{align*}
Again, it should be noted that the representation in Eq.~\ref{eq:likelihood} is essential for a derivation of our model selection algorithm in the following section.

\subsection{PHMM model selection algorithm: FAB-PHMM}\label{sec:FAB-PHMM}
We formalise the model selection problem for PHMM as a maximisation of the model evidence. 
\begin{eqnarray}
\mathcal{M}^* = \argmax_{\mathcal{M}} \ln p(\vec{X}| \mathcal{M})
\end{eqnarray}
where the evidence is given by $p(\vec{X}| \mathcal{M}) = \int \sum_{\vec{Z}}p(\vec{X}, \vec{Z}, \vec{\Pi} | \mathcal{M}) d\vec{\Pi}$. Note that the model size $\mathcal{M} = \{K_M, K_{X}, K_{Y}\}$ is parameterised by the number of hidden states of each state type.
However, the model evidence is difficult to compute; thus, we generally need approximations. In this study, we use FIC as an asymptotically accurate approximation.

FIC has following three appealing properties:
\begin{enumerate}
\item
\textbf{Asymptotic equivalence to marginal likelihood.} Although BIC is a widely used and simple information criterion, it lacks theoretical justification because of the non-regularity of the latent variable models \citep{Watanabe2009}. PHMM is not an exception to this, so the BIC's approximation is invalid for PHMMs. Unlike BIC, FIC is consistent with the marginal likelihood for latent variable models. Practically speaking, \citet{Fujimaki2012a} empirically showed that BIC-HMM tends to choose overly complicated models, while FIC-HMM chooses optimal models more often. 
\item
\textbf{Simultaneous optimisation of model and parameters.} VI is closely related to FIC. Both of them perform similar approximation using variational distribution. One advantage of FIC is it can optimise parameters and models simultaneously. This make FIC-based optimisation computationally more efficient.
\item 
\textbf{Prior free.} Unlike VI, FIC does not require prior distributions because it treats priors as $\mathcal{O}(1)$. Thus, FIC is hyper-parameter tuning free and easier to optimise.
\end{enumerate}

In the following, we will start with the derivation of $FIC_\textit{PHMM}$ (Section~\ref{sec:fic}), then take a lower-bound to derive $FICLB$ (Section~\ref{sec:ficlb}) for optimisation via expectation maximisation (EM). We iteratively optimise the target function $FICLB$ with respect to a variational distribution $q$ and parameters $\vec{\Pi}$ in the E step (Section~\ref{sec:estep}) and M step (Section~\ref{sec:mstep}), respectively. The model $\mathcal{M}$ is tuned via \textit{model pruning} (Section~\ref{sec:model_pruning}).

\subsubsection{FIC} \label{sec:fic}
Let $\vec{\Xi}$ be a set of local (component-dependent) parameters and $\vec{\Theta}$ be a set of global (component-independent) parameters. Additionally, we denote all the parameters as $\vec{\Pi} = \{\vec{\Xi}, \vec{\Theta } \}$ (in the case of PHMM, local parameters $\vec{\Xi} = \{\vec{\beta}, \vec{\phi}\}$ and global parameters $\vec{\Theta} = \{\alpha\}$). \citet{Hayashi2015} have shown that the model evidence can be approximated as an asymptotically accurate information criterion, FIC, which can be expressed as
\begin{align*}
FIC(\mathcal{M}) = \mathbb{E}_{q^{**}} \big[\ln p(\vec{X}, \vec{Z}| \bar{\vec{\Pi}}, \mathcal{M}) - \frac{1}{2} \ln |F_{\bar{\vec{\Xi}}}|\big]
- \frac{1}{2}D_{\vec{\Pi} } \ln N + \mathcal{H}(q^{**}).
\end{align*}
where $D_{\vec{\Pi}}$ is the number of free parameters in $\vec{\Pi}$, $\bar{\vec{\Pi}} = \{ \bar{\vec{\Xi}}, \bar{\vec{\Theta}} \}$ is a maximum joint likelihood estimators (MJLE), $\vec{F}_{\vec{\bar{\Xi}}}$ is the Hessian matrix of $- \ln p(\vec{X}, \vec{Z} | \vec{\Pi})$ with respect to $\vec{\bar{\Xi}}$, $q^{**}(\vec{Z})=p(\vec{Z}|\vec{X}, \mathcal{M})$ is the marginal posterior and  $\mathcal{H}(q^{**})$ is the entropy of $q^{**}$. In FIC, the penalty term is given by the volume of the Fisher information matrix $|F_{\bar{\vec{\Xi}}}|$, which penalises complexity in the model.

Here we derive FIC for PHMM, $FIC_{\textit{PHMM}}$. Since the local parameters $\vec{\Xi} = \{ \vec{\beta}_1, \ldots, \vec{\beta}_K, \vec{\phi}_1, \ldots, \vec{\phi}_K \}$ do not interact with each other, the Fisher information matrix $F_{\vec{\Xi}}$ is a block diagonal matrix whose blocks are $\{ F_{\vec{\beta}_1}, ..., F_{\vec{\beta}_K}, F_{\vec{\phi}_1}, \ldots, F_{\vec{\phi}_K}\}$, thus $\ln |F_{\Xi}| = \sum_k (\ln |F_{\vec{\xi}_k}| + \ln |F_{\vec{\beta}_k}|)$. Here, using the equation (\ref{eq:likelihood}), we can write these Fisher information matrices as
\newcommand{\nabrabeta}{\vec{\nabla}^2_{\vec{\beta}}}
\begin{align*}
F_{\vec{\beta}_k} = \frac{1}{N}
 \sum_{n=1}^N \sum_{t=0}^{T_X } \sum_{u=0}^{T_Y}
{pa(\vec{z}^n_{tu})_k}  \nabrabeta \ln p_k(\vec{z}^n_{tu} | \vec{\beta}_k) 
\end{align*}
and
\newcommand{\nabraphi}{\vec{\nabla}^2_{\vec{\phi}}}
\begin{align*}
F_{\vec{\phi}_k} = \frac{1}{N}
 \sum_{n=1}^N \sum_{t=0}^{T_X } \sum_{u=0}^{T_Y}
{z^n_{tuk}} \nabraphi \ln p(x^n_t, y^n_u | \vec{z}^n_{t u}, \vec{\phi}_k),
\end{align*}
where both $\ln p_k(\vec{z}^n_{tu} | \vec{\beta}_k) $ and $\ln p(x^n_t, y^n_u | \vec{z}^n_{t u}, \vec{\phi}_k) $ are $\mathcal{O}(1)$ with respect to the number of samples $N$. Thus, the penalty term is 
\begin{align*}
&\ln |F_{\vec{\beta}_k}| =  D_{\vec{\beta}_k}\ln \frac{\zeta^{\text{trans}}_k(\vec{Z})}{N} + \mathcal{O}(1) \\
&\ln |F_{\vec{\phi}_k}|  = D_{\vec{\phi}_k}\ln \frac{\zeta^{\text{emit}}_k(\vec{Z})}{N} + \mathcal{O}(1)\\
\text{where} & \quad \begin{cases}
\zeta^{\text{trans}}_k(\vec{Z}) &= 
 \sum_{n=1}^N \sum_{t=0}^{T_X } \sum_{u=0}^{T_Y}
{pa(\vec{z}^n_{tu})_k} \\
&=\sum_{n=1}^N \sum_{t=0, u=0}^{T^n_x, T^n_y} z^n_{t u k} - \sum_{n=1}^N z^n_{T^n_x, T^n_y}.\\
\zeta^{\text{emit}}_k(\vec{Z}) &= \sum_{n=1}^N \sum_{t=0, u=0}^{T^n_x, T^n_y} z^n_{t u k}.
\end{cases}
\end{align*}
The newly introduced symbols $\zeta^{\text{emit}}_k$ and $\zeta^{\text{trains}}_k$ are \textit{effective samples} of, respectively, transition and emission probability for the $k$-th latent variable. The values $D_{\vec{\beta}_k}$ and $D_{\vec{\phi}_k}$ are the dimensionalities of parameters $\vec{\beta}_k$ and  $\vec{\phi}_k$, respectively.

Finaly, ignoring the $\mathcal{O}(1)$ term, we derive FIC for PHMM as
\begin{align*}
FIC_{\textit{PHMM}}(\mathcal{M}) = 
\mathbb{E}_{q^{**}}\bigg[ \ln p(\vec{X}, \vec{Z} | \bar{\vec{\Pi}})  
 - \sum_{k = 1}^K \frac{D_{\beta_k}}{2}\ln \frac{\zeta^{\text{trans}}_k(\vec{Z})}{N}
- \sum_{k =1} ^K \frac{D_{\phi_k}}{2}\ln \frac{\zeta^{\text{emit}}_k(\vec{Z}) }{N} \bigg] - \frac{D_\Pi}{2} \log N + \mathcal{H}(q^{**} ).
\end{align*}

The penalty terms  are now sums of parameter dimensionality weighted by the corresponding effective samples. For example, the dimensionality of the $k$-th emission probability $D_{\phi_k}$ is weighted by $\zeta^{\text{emit}}_k(\vec{Z})$. When the effective sample of the $k$-th component is small, the penalty term for the $k$-th latent variable also becomes small. In this case, $\vec{Z}$ is \textit{degenerate} and we can safely prune the $k$-th latent component. This model pruning is further discussed in Section \ref{sec:model_pruning}.

\subsubsection{FIC Lower-bound}
\label{sec:ficlb}
{
We employ an EM algorithm to optimise the parameters. To make the EM algorithm tractable, we further take the lower bound of $FIC_\text{PHMM}$ and derive $FICLB$. We use three approximations to construct the lower bound. (1) Since the MJLEs $\bar{\vec{\Pi}}$ is unavailable in practise, we replace it by the arbitrary parameter $\vec{\Pi}$, which is optimised in the M step. (2) Instead of the marginal posterior $q^{**}$, we use a variational distribution $q$, which is optimised in the E step. (3) We take a lower bound of the negative logarithm as $- \log (\sum_{ntu} z^n_{tuk}) \geq - L( \sum_{ntu} z^n_{tuk}, \sum_{ntu} \tilde{q}(z^n_{tuk}))$, where $L$ is linear approximation of the logarithm function $L(a, b) = \log b + (a - b) / b$ and $\tilde{q}$ is any distribution over $\vec{Z}$. During the optimisation procedure, $\tilde{q}$ is set to be the variational distribution $q$ of the previous time step. Using these approximations, we now get the lower bound}
{
\begin{align*}
FIC_{\text{PHMM}} (\mathcal{M}) &\geq FICLB(\mathcal{M}, q, \tilde{q}, \vec{\Pi}) \\
&= 
\mathbb{E}_{q} \bigg[ \log p(\vec{X}, \vec{Z} | \vec{\Pi}) 
+\sum_{n=1}^N \sum_{t=0, u=0}^{T^n_x, T^n_y} z^n_{t u k} \log \delta_{t u k} \bigg]
- \frac{D_\alpha}{2} \log N  
\nonumber\\ 
&\qquad
- \sum_{k=1}^K \frac{D_{\beta_k}}{2}\log \big(\sum_{n, t, u} 
\zeta_k^{\text{trans}}(\tilde{\vec{Z}})- 1 \big)
- \sum_{k=1}^K \frac{D_{\phi_k}}{2}\log \big(\zeta_k^{\text{emit}}(\tilde{\vec{Z}}) - 1 \big) + \mathcal{H}(q) \quad \nonumber\\
&\text{where}\quad \delta_{t u k} = \left \{
\begin{aligned}
&\exp \bigg( - \frac{D_{\phi_k}}{2 \zeta_k^{\text{emit}}(\tilde{\vec{Z}}) } \bigg) \enskip \text{if} \enskip t=T_X \enskip \text{and} \enskip u = T_Y 
\\
& \exp \bigg( - \frac{D_{\phi_k}}{2 \zeta_k^{\text{emit}}(\tilde{\vec{Z}}) } - \frac{D_{\beta_k}}{2 \zeta_k^{\text{trans}}(\tilde{\vec{Z}}) } \bigg) \enskip  \text{otherwise}.
\end{aligned} \right.
\label{eq:ficlb}
\end{align*}
}
Here, we introduced the auxiliary variable $\tilde{\vec{Z}} = \{\tilde{q} (z^n_{t u k})\}_{t,u,k,n}$ for simplicity.
The full algorithm including \textit{model pruning} (the model selection mechanism) is explained in Section \ref{sec:model_pruning}.

\subsubsection{E-step updates} \label{sec:estep}
We need to obtain the distribution $q^*$ that maximises $FICLB$ (see Algorithm \ref{alg:fab} for details). This can be done using a modified forward--backward algorithm as follows.
\begin{align}
f^n_{tuk} &= 
\begin{cases}
 0  \qquad \text{if $t<0$ or $u<0$ or $(t,u)=(0,0)$} \\
 \alpha_k \psi^n_{tuk} \delta_{tuk} \qquad \text{if initial position} \\
 \psi^n_{tuk} \delta_{tuk} \sum_{j=1}^K f^n_{t-d_{xk}, u-d_{yk}, k } \beta_{j,k} \qquad \text{otherwise}
\end{cases} \\
b^n_{tuk} &= 
\begin{cases}
    0  \qquad \text{if $t > T^n_x$ or $u > T^n_y$} \\
    1 \qquad\text{if $t = T^n_x$ and $u = T^n_y$}\\
    \sum_{l=1}^K 
 \psi^n_{tul} \delta_{tul} b^n_{t+d_{xl}, u+d_{yl}, l } \beta_{k,l} \qquad \text{otherwise}
\end{cases}
\end{align}
Using these forward--backward variables, the optimal variational distribution $q^*$ is obtained as follows.
\begin{align}
q^*(z^n_{tuk}) 
&= \frac{f^n_{tuk} b^n_{tuk}}{\sum_l f^n_{T^n_x, T^n_y, l}} \\
q^*(z^n_{t-d_{xk}, u - d_{xy}, j}, z^n_{tuk}) 
&= \frac{f^n_{t-d_{xk}, u - d_{xy}, j}\beta_{jk} \phi^n_{tuk} b^n_{tuk}}{ \sum_l f^n_{T^n_x, T^n_y, l} }
\end{align}

\subsubsection{M-step updates}\label{sec:mstep}
Now, we want to find the $\vec{\Pi}$ that maximises $FICLB$ for fixed $q$ (see Algorithm \ref{alg:fab} for details). For those parameters, we have the update function
\begin{align}
\alpha_k &\propto \sum_{n} q(z^n_{d_{xk}, d_{yk}, k}) \\
\beta_{jk} &\propto \sum_{n,t,u} q(z^n_{t-d_{xk}, u - d_{xy}, j}, z^n_{tuk}) \\
\phi_k(x, y) &\propto 
\begin{cases}
\sum_{ntu} q(z^n_{tuk}) \mathbb{I}(x=x_t \land y=y_u) \qquad\text{if $\mathcal{S}(k) = M$} \\
\sum_{ntu} q(z^n_{tuk}) \mathbb{I}(x=x_t ) \qquad\text{if $\mathcal{S}(k) = X$} \\
\sum_{ntu} q(z^n_{tuk}) \mathbb{I}(y=y_u ) \qquad\text{if $\mathcal{S}(k) = Y$}
\end{cases}.\nonumber
\end{align}
For calculation of $\beta_{jk}$, out-of-range indexing is treated as zero, that is $q(z^n_{d_{xk}, d_{yk}, k}) = 0$ if $t-d_{xk} < 0$ or $u-d_{yk} < 0$.

\subsubsection{Pruning degenerated components}\label{sec:model_pruning}
In contrast to variational inference, the FAB algorithm enables simultaneous optimisation of a model $\mathcal{M}$ and its parameters $\vec{\Pi}$ via \textit{model pruning} \citep{Hayashi2015}. Let us call $\vec{Z}$ \textit{degenerated} when there exists an equivalent likelihood for a smaller model $\tilde{\mathcal{M}}$, that is $p(\vec{X}, \vec{Z}| \vec{\Pi}, \mathcal{M}) = p(\vec{X}, \tilde{\vec{Z}}| \tilde{\vec{\Pi}}, \tilde{\mathcal{M}}), $ where $\mathcal{M} > \tilde{\mathcal{M}}$. In such cases, $\mathcal{M}$ is overcomplete so we can transform the model $(\vec{Z}, \vec{\Pi}) \rightarrow (\tilde{\vec{Z}}, \tilde{\vec{\Pi}}) $ to obtain a new smaller and equivalent model. This transformation is called \textit{model pruning}. In the case of FAB-PHMM, we can prune the components $k$ with \textit{effective samples} $\sum_{ntu} q(z^n_{tuk}) / N$ that are beneath some threshold $\epsilon$. Starting from a sufficiently large model, we can prune redundant components while optimising parameters. This algorithm is shown in Algorithm~\ref{alg:fab}.
 
 We observed that this model pruning algorithm sometimes fails by being captured within poor local optima. In such cases, degenerated components are not pruned. To avoid this problem, we incorporate \textit{greedy pruning} (Algorithm~\ref{alg:greedy_fab}). 
 When the algorithm converged, we append the current model $(\vec{\Pi}, \mathcal{M}, q)$ to model candidates. Then, delete the component with the fewest effective samples (greedy\_pruning) and restart the algorithm. 
 After finding the smallest possible model $(K_M, K_X, K_Y) = (1,1,1)$, we choose the model with the largest FIC from among the model candidates. 
 
{ 
\subsubsection{Computational complexity}
For each iteration in Algorithm \ref{alg:fab}, the computational complexity is $\mathcal{O}(N \max(T_X^n) \max(T_Y^n) K^2)$ for the E and M steps, and $\mathcal{O}(N \max(T_X^n) \max(T_Y^n) K)$ for model pruning. Therefore, the overall complexity for each step is $\mathcal{O}(N \max(T_X^n) \max(T_Y^n) K^2)$. Note that this complexity is exactly the same as { ordinary} parameter learning for PHMM using the Baum--Welch algorithm \citep{Durbin1998}.
}

%


%
\vspace{2em}
\begin{algorithm}
\caption{The FABPHMM algorithm}
\label{alg:fab}
\begin{algorithmic}
\State \textbf{Input:} data $\vec{X}$, initial model $\mathcal{M} = (K_M, K_X, K_Y)$, initial variational distribution $q$, 
initial parameter $\vec{\Pi}$, stopping threshold $\eta$ and pruning threshold $\epsilon$
{\State $FICLB_\text{prev} = \infty$
\Loop 
\State $\tilde{q} \leftarrow q$ 
\State $q \leftarrow \argmax_{q} FICLB(\mathcal{M}, q, \tilde{q}, \vec{\Pi})$ \Comment{E-step}
\ForAll{$k$ that satisfy $\sum_{n,t,u} q(z^n_{tuk}) \leq \epsilon$} 
\State delete the $k$-th hidden state of model $\mathcal{M}$ \Comment{Pruning}
\EndFor
\State $\vec{\Pi} \leftarrow \argmax_{\tilde{\vec{\Pi}}} FICLB(\mathcal{M}, q, \tilde{q},\tilde{\vec{\Pi}})$ \Comment{M-step}
\If{$|FICLB(\mathcal{M}, q, \tilde{q},\vec{\Pi}) - FICLB_\text{prev}| < \eta$}
    \State end loop
\EndIf
\State $FICLB_\text{prev} \leftarrow FICLB(\mathcal{M}, q, \tilde{q},\tilde{\vec{\Pi}})$
\EndLoop
}
\end{algorithmic}
\end{algorithm}

\vspace{-2em}
\begin{algorithm}
\caption{The greedy FABPHMM algorithm}
\label{alg:greedy_fab}
\begin{algorithmic}
\State \textbf{Input:} data $\vec{X}$, initial model $\mathcal{M} = (K_M, K_X, K_Y)$, initial variational distribution $q$, 
initial parameter $\vec{\Pi}$, stopping threshold $\eta$ and pruning threshold $\epsilon$

\State Initialize $candidates$ with an empty list
\Loop
\State $(\bar{\vec{\Pi}}, \bar{\mathcal{M}}, \bar{q})$ $\leftarrow$ FABPHMM-algorithm($\vec{X}, \vec{\Pi}, \mathcal{M}, q, \eta, \epsilon$)   \Comment{Alg. \ref{alg:fab}}
\State append  $(\bar{\vec{\Pi}}, \bar{\mathcal{M}}, \bar{q})$ to $candidates$
\If{$\bar{\mathcal{M}} = (1,1,1)$}
    \State end loop
\EndIf
\State ($\vec{\Pi}, \mathcal{M}, q$) $\leftarrow $ greedy\_pruning($\bar{\vec{\Pi}}, \bar{\mathcal{M}}, \bar{q}$) \Comment{Delete the component with fewest effective samples}
\EndLoop
\State choose model with highest FIC from $candidates$
\end{algorithmic}
\end{algorithm}

\section{Experiments}\label{sec:exp}

We performed three types of experiments to answer the following three questions. First, how accurate is the proposed method in selecting the optimal model (Section~\ref{sec:exp_model_selection})? Next, how much does the proposed method contribute to alignment accuracy relative to fixed-size models (Section~\ref{sec:exp_align_accuracy})? Finally, what kinds of models are selected when we train our proposed method against real DNA data (Section~\ref{sec:exp_real})?

In this study, because of the high computational cost of parameter learning, we concentrated on short alignments (up to 200 bp). Additionally, we considered only global alignments. Extensions for the alignment of longer sequences and local alignment will be discussed later (in the Discussion section). Moreover, following previous research \citep{Lunter2008,pmid19478997}, we here concentrate on models with a single match state and multiple insertion states, although our method is potentially applicable to multiple match states (cf. Figure~\ref{fig:hidden_states} and the Discussion section).
For all the experiments, we set the stopping threshold $\eta = 10^{-5}$ and pruning threshold $\epsilon = 10^{-4}$. 

\subsection{Model selection capability}\label{sec:exp_model_selection}

We first investigated the model selection capability of the proposed method. We used synthetic data because the \textit{true model} is not available for real dataset. We defined parameters of PHMMs of different sizes manually and generated alignments from them. The true model size $\mathcal{M}_{\text{true}} = (K^*_M, K^*_X, K^*_Y)$ and their names are shown in Table \ref{tab:art_models}. From the manual models, we generated $N$ alignments of fixed length $100$. After removing the gaps from each alignment, we fed the sequences to Algorithm \ref{alg:greedy_fab} and estimated the optimal model only from the data (pairs of sequences). We then determined if it can recover the true model. We set the initial model size to be $(K_M, K_X, K_Y)=(1, 10, 10)$. We ran experiments for sample sizes $N=100,200,300,400,500,600,700,800,900,1000$. 
Parameters for each model can be found in supplementary section S3 and Figures S1--S7.


Table~\ref{tab:model_selection_multi} shows the fraction of models that were correctly predicted -- prediction is correct when the true model size and predicted model size are exactly the same. 
As the number of samples $N$ grew, the approximation became more precise. Also, some models tended to require more samples for precise model selection. With sufficient samples, FAB-PHMM successfully recovered models almost perfectly except in some cases (i.e. in experiments (\texttt{med},$N=1000$) and (\texttt{small},$N=1000$)). Even in these cases, the inaccurately selected models have smaller FIC values than accurately selected models, and we assumed the algorithm was trapped in a poor local optimum. However, we can easily avoid this problem by using multiple runs and choosing the model with the largest FIC. Indeed, when we picked such a model out of 10 replicates generated with a different random seed, the predictions were always correct when the sample size was greater than or equal to 700 (see bold-face text in Table \ref{tab:model_selection_multi}; the bold-face font indicates that the model with the highest FIC value successfully predicted the correct model).

Note that this experimental setting is much simpler than the real setting, where the true model is not in the PHMM class (i.e. the real DNA sequences are not PHMM generated). However, we assume that FAB-PHMM can select the optimal model, in the sense of choosing the closest possible models.

\begin{table}[tb]
\caption{Model sizes of hand-crafted models. The 1st, 2nd and 3rd values in each triplet are the numbers of match states ($K_M$), $X$-insertion states ($K_X$) and $Y$-insertion states ($K_Y$), respectively. Model names including the term "imb" indicate an imbalanced model with unequal values of $K_X$ and $K_Y$.  See Supplementary Figures S1--S7 for details of the parameters.\label{tab:art_models}}  
\begin{center}
\begin{tabular}{@{}ccccccc@{}}
\toprule 
\texttt{small} & \texttt{med} & \texttt{large} & \texttt{imb} & \texttt{imb\_large} &  \texttt{huge} & \texttt{imb\_huge}\\ 
\midrule 
(1, 1, 1) & (1, 2, 2) & (1, 4, 4) & (1, 2, 1) & (1, 4, 2) & (1, 6, 6) & (1, 6, 3)\\ 
\botrule 
\end{tabular}
\end{center}
\end{table}


\subsection{Alignment accuracy}\label{sec:exp_align_accuracy}

We also explored the alignment accuracy of the proposed method. Since it is difficult to obtain true genome alignments, we re-used the generated data from the manual models in the model selection experiment. For alignment accuracy assessments, we had five alignment datasets for each different model size. We first trained multiple models on each of those datasets and then performed alignments. For training, we used PHMM (with a fixed model size) of those eight different model sizes with random parameter initializations in addition to FAB-PHMM, which automatically chose the optimal model size. In total, we used nine PHMMs, including one with the true model sizes as well as one FAB-PHMM. In order to avoid poor local optima, we ran five trainings for each setting and selected the model with the best score (FIC for FAB-PHMM and likelihood for PHMM). 

As a measure of performance in terms of accuracy, we used the f1 score, which is the harmonic mean of precision and recall:
\begin{align}
    \text{f1} &= 2\frac{\text{precision} \cdot \text{recall}}{\text{precision} + \text{recall}} \\
    \text{precision} &= \frac{\text{\# correctly inferred positions}}{\text{\# inferred positions}} \\
    \text{recall} &= \frac{\text{\# correctly inferred positions}}{\text{\# true positions}}.
\end{align}
For example, when the true alignment is $\bfrac{x_1 x_2 x_3 x_4}{- \ y_1 y_2 y_3}$ and the inferred alignment is $\bfrac{x_1 x_2 x_3 x_4}{y_1 \ - y_2 y_3 }$, the true and inferred positions are $\{(x_2, y_1), (x_3, y_2), (x_4, y_3)\}$ and $\{(x_1, y_1), (x_3, y_2), (x_4, y_3)\}$, respectively. The correctly inferred position is simply the intersection of the true positions and inferred positions: $\{(x_3, y_2), (x_4, y_3)\}$. In this case, precision is 2/3 and recall is 2/3; thus, the f1 score is 2/3.

We report the result of the number of sequences where $N=1000$ in Table \ref{tab:f1_pos}. For every dataset, the proposed method performed on par with the true model while all the other fixed-size models performed relatively poorly in some cases, for example for the \texttt{large} model dataset experiment, FAB-PHMM and the \texttt{large} model (i.e. the same model size as the one that produced dataset) performed better than others, whereas the \texttt{large} model performed relatively poorly for smaller datasets. 

Although this alignment accuracy measure is widely used (e.g. in \cite{Rivas2015}), this approach only considers aligned bases that correspond to match states and ignores all those corresponding to insertion states. For this reason, we also evaluated the insertion counterparts of the f1 score. (See Supplementary Section~S1 for further detail.)

In addition to assessing alignment accuracy, we also calculated the perplexity of each trained model in order to show how well the models explain the data. Refer to Supplementary Section~S2 for further detail.

\begin{table}[tb]
\caption{Fraction of precise model selections from multiple estimations for data produced with different model sizes. We ran the model selection algorithm 10 times for each combination of sample number and model size (shown in Table \ref{tab:art_models}) with random initial parameters. Each table entry indicates the fraction of runs in which the correct optimal model was selected. The bold-face font indicates precise prediction of the model size, that is when the method selected the model with the largest FIC correctly. \label{tab:model_selection_multi}} 
\begin{center}{
\begin{tabular}{@{}rccccccc@{}} 
\toprule 
\multirow{ 2}{*}{\# samples}& \multicolumn{7}{c}{precise model selection for different model sizes} \\ 
\cline{2-8} 
 & \texttt{small} & \texttt{med} & \texttt{large} & \texttt{imb} & \texttt{imb\_large} & \texttt{huge} & \texttt{imb\_huge}\\ 
\midrule 
100 & \textbf{ 10/10 } & 0/10 & 0/10 & \textbf{ 10/10 } & 0/10 & 0/10 & 0/10\\ 
200 & \textbf{ 10/10 } & 0/10 & 0/10 & 9/10 & 0/10 & 0/10 & 0/10\\ 
300 & \textbf{ 10/10 } & 0/10 & 0/10 & \textbf{ 4/10 } & 7/10 & 0/10 & 0/10\\ 
400 & \textbf{ 10/10 } & 0/10 & 0/10 & \textbf{ 9/10 } & \textbf{ 10/10 } & 0/10 & \textbf{ 10/10 }\\ 
500 & \textbf{ 10/10 } & 0/10 & 0/10 & \textbf{ 7/10 } & \textbf{ 9/10 } & 0/10 & 5/10\\ 
600 & \textbf{ 10/10 } & \textbf{ 5/10 } & 0/10 & \textbf{ 9/10 } & \textbf{ 10/10 } & \textbf{ 5/10 } & \textbf{ 9/10 }\\ 
700 & \textbf{ 10/10 } & \textbf{ 10/10 } & 3/10 & \textbf{ 5/10 } & \textbf{ 10/10 } & \textbf{ 10/10 } & \textbf{ 10/10 }\\ 
800 & \textbf{ 10/10 } & \textbf{ 10/10 } & \textbf{ 10/10 } & \textbf{ 7/10 } & \textbf{ 10/10 } & \textbf{ 10/10 } & \textbf{ 10/10 }\\ 
900 & \textbf{ 9/10 } & \textbf{ 10/10 } & \textbf{ 10/10 } & \textbf{ 8/10 } & \textbf{ 10/10 } & \textbf{ 10/10 } & \textbf{ 10/10 }\\ 
1000 & \textbf{ 9/10 } & \textbf{ 8/10 } & \textbf{ 10/10 } & \textbf{ 7/10 } & \textbf{ 10/10 } & \textbf{ 10/10 } & \textbf{ 10/10 }\\ 
\botrule 
\end{tabular}
}
{}
\end{center}
\end{table}
\begin{table*}[tb]
\caption{Alignment f1 score. For each model of data generation (i.e. the simulated models shown in Table \ref{tab:art_models}), we trained models of fixed sizes including the true model and proposed model (\texttt{fab}) before the f1 score evaluation. The italic and bold values indicate the result of training with the true model size and the best score obtained without the true model size, respectively. See Table~\ref{tab:art_models} for the details of the models.
\label{tab:f1_pos}} 
\centerline{
\begin{tabular}{@{}lcccccccc@{}} 
\toprule 
              & \multicolumn{8}{c}{Training models} \\
Simulated models & \texttt{small} & \texttt{med} & \texttt{large} & \texttt{imb} & \texttt{imb\_large} & \texttt{huge} & \texttt{imb\_huge} & \texttt{fab}\\ 
\midrule 
\texttt{small} & \textit{ 0.9286 } & 0.9286 & 0.9255 & 0.9283 & 0.9281 & 0.9247 & 0.9262 & \textbf{ 0.9287 }\\ 
\texttt{med} & 0.8337 & \textit{ 0.8327 } & 0.8318 & \textbf{ 0.8342 } & 0.8324 & 0.8258 & 0.8279 & 0.8328\\ 
\texttt{large} & 0.8196 & 0.8255 & \textit{ 0.8309 } & 0.8238 & 0.8278 & 0.8277 & 0.8313 & \textbf{ 0.8315 }\\ 
\texttt{imb} & 0.8882 & 0.8963 & 0.8927 & \textit{ 0.8965 } & 0.8925 & 0.8919 & 0.8928 & \textbf{ 0.8968 }\\ 
\texttt{imb\_large} & 0.8602 & \textbf{ 0.8694 } & 0.8681 & 0.8659 & \textit{ 0.8686 } & 0.8654 & 0.8678 & 0.8688\\ 
\texttt{huge} & 0.9419 & 0.9473 & 0.9506 & 0.9449 & 0.9490 & \textit{ 0.9513 } & 0.9510 & \textbf{ 0.9515 }\\ 
\texttt{imb\_huge} & 0.9681 & 0.9688 & \textbf{ 0.9717 } & 0.9690 & 0.9715 & \textbf{ 0.9717 } & \textit{ 0.9719 } & \textbf{ 0.9717 }\\ 
\midrule
average & 0.9035 & 0.9066 & 0.9071 & 0.9061 & 0.9069 & 0.9052 & 0.9068 & \textbf{ 0.9083 }\\ 
\botrule 
\end{tabular}
}
{}
\end{table*}



%
\subsection{Model selection from real DNA sequences}\label{sec:exp_real}

We used real DNA data to explore the resulting models selected by FAB-PHMM. Because this study concentrates on global alignments of short sequences, proper analyses require homologous DNA sequence pairs. For this reason, we used locally-aligned DNA data produced by \citet{Frith2015}\footnote{https://zenodo.org/record/17436\#\#.WA3REpOLQYM} and multiple-aligned DNA data produced by MULTIZ\footnote{http://hgdownload-test.cse.ucsc.edu/goldenPath/hg38/multiz20way/}.

\subsubsection{LAST dataset}

The LAST dataset \citep{Frith2015} contains pairwise alignments of human sequences to those of four other species (dog, orangutan, mouse and chimpanzee). For our purposes, we selected the alignments with lengths between 100 and 200. Additionally, we removed the alignments with a "missmap" probability (an alignment ambiguity measure provided with the dataset) of less than $10^{-5}$ because we wanted to use only highly-reliable homologous pairs. This resulted in 1640 to 94,904 remaining alignments for each of the alignments between the human sequences and those of the four species. We extracted overlapping regions from across all four species and cropped alignments to include only those overlapping regions. Then, we randomly sampled 1000 alignments and removed gaps from them in order to input them into FAB-PHMM. We ran 10 trainings with different random seeds (which determine the initial parameters in PHMM) and selected the best model with the highest FIC value. We set the initial model size to be $(K_M, K_X, K_Y)=(1, 12, 12)$. 

\begin{figure}
\centerline{\includegraphics[width=1.0 \columnwidth]{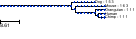}}
\vskip -5em
  \caption[]{Trained model sizes for the LAST dataset shown on a phylogenetic tree generated using phyloT and the ETE Toolkit. The model sizes are to the right of the species names, for example ``Mouse - 1 6 3'' means the selected model for the human--mouse alignment is $(K_M, K_X, K_Y) = (1,6,3)$.}\label{fig:last_tree}
\end{figure}

Figure \ref{fig:last_tree} shows the selected model size for each species. For species more closely related to humans (i.e. orangutan and chimpanzee), the algorithm selected a similar model and vice versa for species more distantly related to humans (i.e. dog and mouse). 
Specifically, in the case of chimpanzee and orangutan alignments, the simplest model $(K_M,K_X,K_Y)=(1,1,1)$ was chosen, while many more X- and Y-insertion states were estimated in the case of dog and mouse alignments. 
Additionally, it was observed that more X-insertion states were predicted than Y-insertion states for dog and mouse alignments.

\begin{figure}[tb] 
\begin{tabular}{cc}
 (a) initial/transition probabilities & (b) emission probabilities\\
 \includegraphics[width=0.44\linewidth]{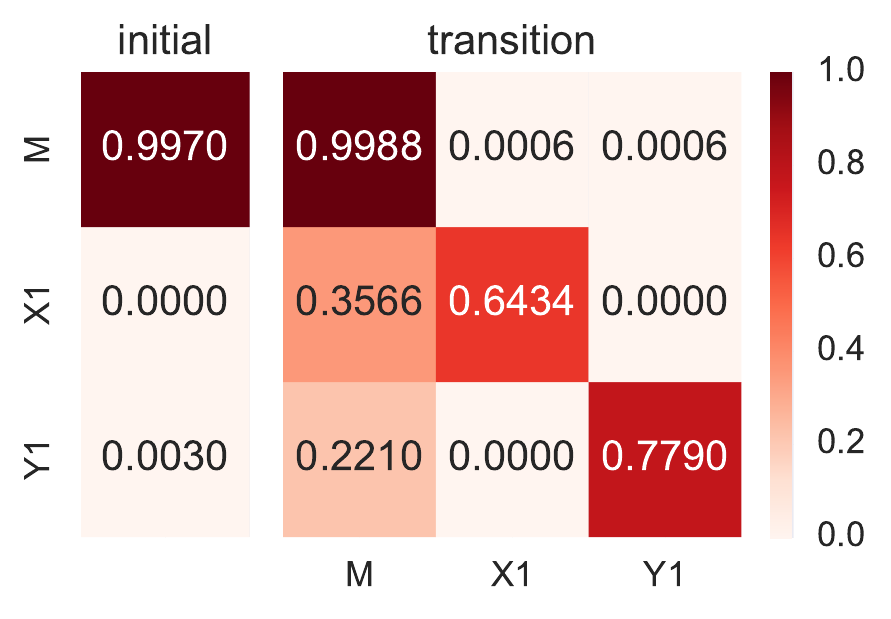}
 & \includegraphics[width=0.48\linewidth]{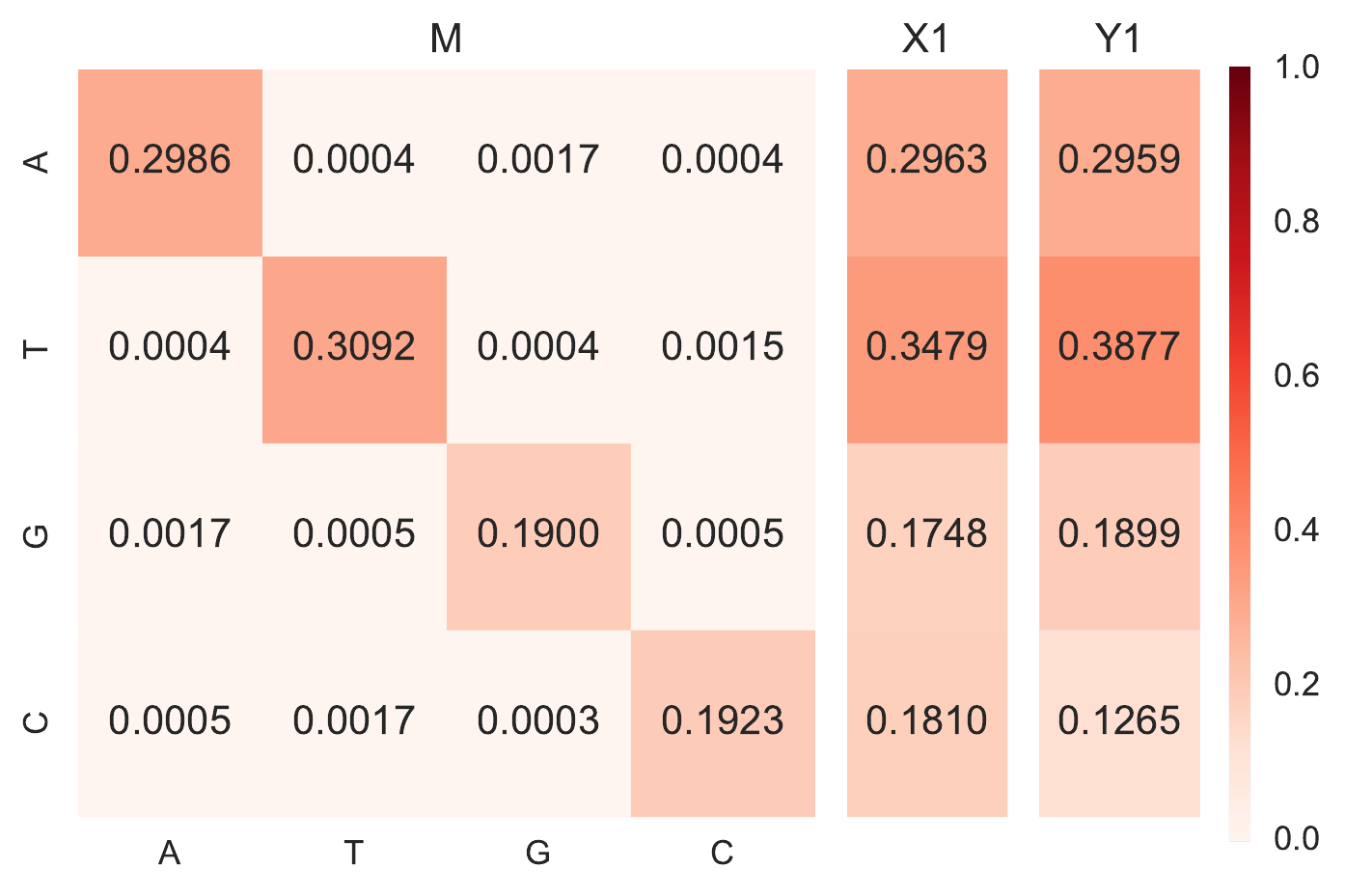}
\end{tabular}

  \caption{Trained PHMM for human--chimpanzee alignments for sequences from the LAST dataset using the proposed method. The resulting model is the simplest one, $(K_M, K_X, K_Y) = (1, 1, 1)$. (a) Trained initial and transition probabilities and (b) emission probabilities.}
  \label{fig:params_chimp}
\end{figure}

\begin{figure}[tb]
\centering
\begin{tabular}{c}
(a) initial/transition probabilities\\
  \includegraphics[width=.8 \linewidth]{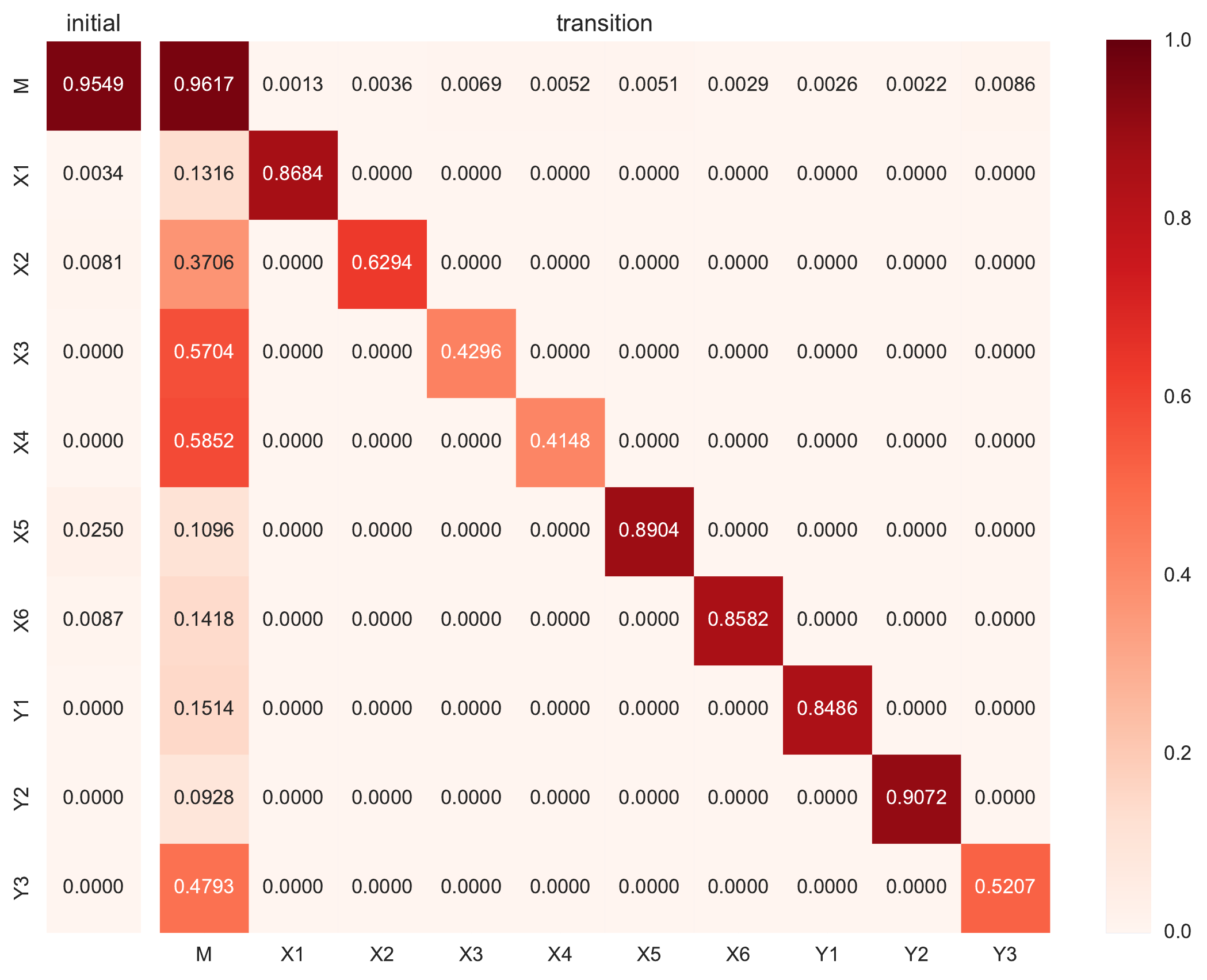}\\
(b) emission probabilities\\
  \includegraphics[width=.8 \linewidth]{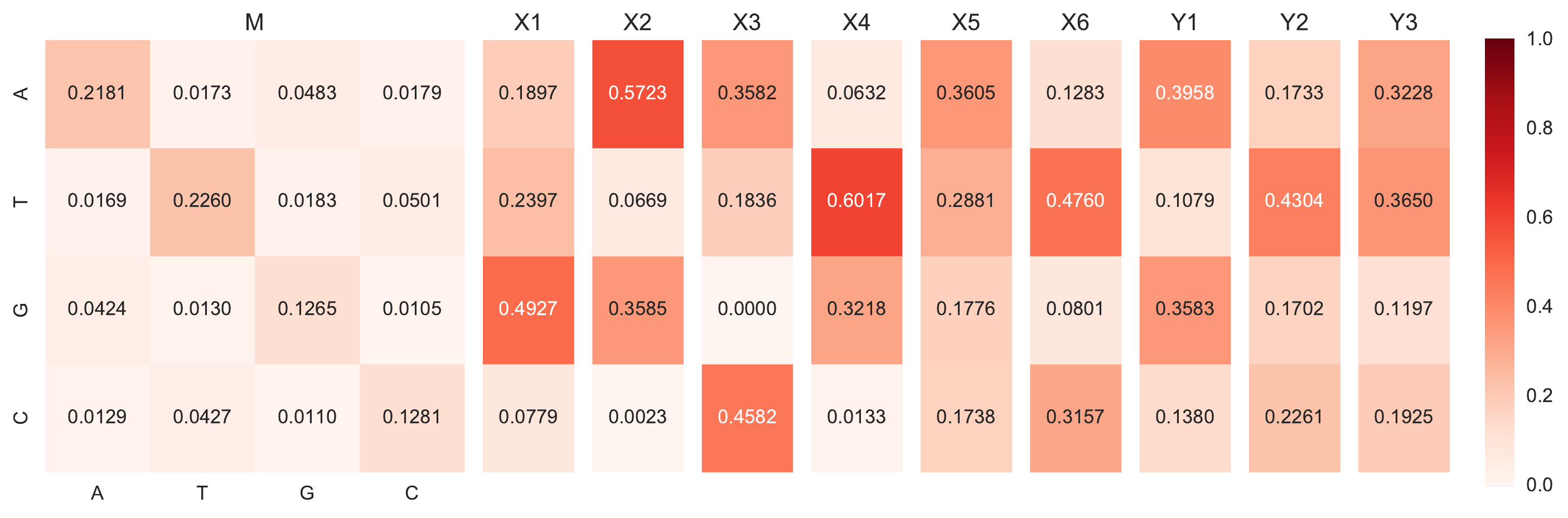}
\end{tabular}

  \caption{Trained PHMM and its parameters for alignments between human and mouse sequences. Panel (a) shows $X1$--$X5$ and $Y1$--$Y3$ have five $X$-insertion states and three $Y$-insertion states, respectively, where each value is an estimated initial/transition probabilities. For example, the value of $0.4793$ in cell $(M,Y3)$ is equal to the transition probability from $Y3$ to $M$. Panel (b) shows emission probabilities of each hidden states. For example, the value 0.0173 in cell $(T, A)$ in the left most panel is equal to the probability of a match state emitting a nucleotide pair $(x=A, y=T)$. } 
  \label{fig:params_mouse}
\end{figure}

Figure~\ref{fig:params_chimp}, Figure~\ref{fig:params_mouse}, Supplementary Figure~S8 and Supplementary Figure~S9 provide the detailed parameters of the selected models from human--chimpanzee, human--mouse, human--orangutan and human--dog sequence alignments, respectively.  
Overall, all the trained substitution matrixes are almost symmetric.
We have the following observations about the human--mouse alignment: two $X$-insertion states, X3 and X4, with similar and relatively high transition probabilities from the match state (M) were predicted, and X3 had higher emission probabilities of A and C while X4 has higher emission probabilities of T and G. 
In contrast, the human--dog sequence alignment provides the following observations: (1) the self-transition probabilities of X1, X3 and X5 were similar (about 0.8), though the emission probability of X1 had an almost uniform distribution while those of X3 and X5 were skewed with different profiles; (2) the self-transition probability of X2 and X4 were smaller (about 0.4) than those of the others, meaning the states corresponded to shorter gaps; (3) we obtained two long insertion states, Y1 and Y3, and one short insertion state, Y2, while Y2 and Y3 had similar emission profiles.

%
%

\subsubsection{MULTIZ dataset}
We also used the multiz20way dataset, which consists of multiple alignments of 19 different species' genome assemblies to the human genome. As with the previous experiment, we randomly select 1000 alignments of human sequences to sequences from each of the 19 species where each was restricted to have lengths of 90 to 110 bp. These alignments were then used as a training set in FAB-PHMM. We set the initial model size to be $(K_M, K_X, K_Y)=(1, 12, 12)$.

\begin{figure}
\centerline{\includegraphics[width=0.85\columnwidth]{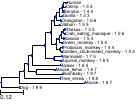}}
    \vspace{-4em}
  \caption[]{Trained model sizes for the MULTIZ dataset using the proposed method with an inferred phylogenetic tree. See the caption of Figure~\ref{fig:last_tree}.}
  \label{fig:multiz_tree}
\end{figure}

The selected model sizes are shown on a phylogenetic tree in Figure \ref{fig:multiz_tree}. In general, we can see species that are more distantly related to humans tend to have alignments with humans that are described by large models, for example the human--bushbaby alignment model has a model size of $(1,9,7)$ and the human--dog alignment model is $(1,9,9)$. However, for species that are more closely related to humans, somewhat random model sizes were selected to describe alignments, for example the human--gorilla alignment model size is $(1,3,3)$ whereas the human--chimpanzee model (a closer species' alignment model) has a model size of $(1,5,4)$. 
%

\section{Discussion}
\label{sec:dis}


Although our proposed method may be potentially adapted to have multiple match states (cf. Figure~\ref{fig:hidden_states}), we have concentrated on a method with a single match state and multiple ($X$- and $Y$-) insertion states. This is because {some} previous studies focused on only multiple insertion states, and {learning} multiple match states { in addition to multiple insertion states} would lead to more complex models that are not interpretable. {(Also, the use of multiple match states incurs more computational cost due to the increase in trained parameters.)}
Still, it might be interesting to investigate multiple match states for further improvement of alignment accuracy or making novel inferences about sequence evolution{, because multiple match states could correspond to the substitution rate of regions}.


Our experiments in Section~\ref{sec:exp_align_accuracy} show that models trained by our proposed method achieved better alignment accuracy than other models, but the improvement was only marginal. We assume this is because 
maximising model evidence not always result in maximising alignment accuracy: higher model evidence might contribute to better sequence modelling, but it might not affect alignment accuracy metric.
This result is consistent with previous research by \citet{Lunter2008} in which the authors indicated that modifications of insertion states can result in only small improvements in alignment accuracy. 




In our analyses using real data (Section~\ref{sec:exp_real}), we utilised limited datasets owing to the high computational cost of our proposed method. Even with these limited datasets, we observed that much more complex models than traditionally used models are selected as optimal. This result implies a possibility that much more complex probabilistic structures exist behind the probabilistic alignments than previously believed. 

As well as improving alignment accuracy, the selected model structure may provide interesting insight about biological sequences because the  selected probabilistic model structure contains latent information of input data.
In other words, the selected model may provide insights into the biological functions of sequences. 
Similarly, decoded hidden states of PHMMs may reveal (e.g. in DNA alignments) that regions with different hidden states correspond to different functions, such as exon versus intron sequences,
non-coding RNAs and regulatory elements. This information can be useful for inferring novel biological insights from sequences.

For the best model selection for real data, however, a comprehensive dataset (including e.g. coding, non-coding and repetitive regions) is required because the trained model structures depend on input data. Indeed, our analyses led to different model selections based on the LAST and MULTIZ datasets. This could be because the homologous pairs taken for the LAST dataset were more similar than those taken for the MULTIZ dataset due to the protocols of generating input sequences (cf. Section~\ref{sec:exp_real}). 

In our future work, we will utilise larger and unbiased datasets in order to select more reliable models. The main bottleneck is the high computational cost of our algorithm. To address this, we will attempt to accelerate the training process, such as by parallelization of the algorithms, stochastic optimisation \citep{Robbins1951, hoffman13a, liu15} and seed-extension heuristics in forward-and-backward algorithms \citep{pmid28039163}




In this study, we focused on genomic DNA sequences, but our method is applicable to RNA or protein sequences as well. The number of characters in protein sequences greatly exceeds those in DNA sequences, which would lead to more complex models that may provide interesting biological insights. These applications will be included in our future research as well.

\section{Conclusion}

In this study, we proposed a novel method to develop PHMMs based on FIC and demonstrated the model selection capability of the proposed model using a synthetic dataset. We believe this is the first study that focuses on model selection of PHMM. On the same synthetic dataset, we observed slight improvement of evaluation metrics of sequence alignments. Additionally, we conducted experiments on real DNA sequences and found that they are best handled with a more complex probabilistic structure than the ones that have been traditionally used for pairwise alignment of these species. This result 
 implies a possibility that
more complex probabilistic structures exist behind probabilistic alignments than previously believed.


\section*{Acknowledgement}
The authors are grateful to Kohei Hayashi (AIST), Tsukasa Fukunaga (Waseda University) and members of the Hamada Laboratory at Waseda University.
Computation for this study was partially performed on the NIG supercomputer at ROIS National Institute of Genetics.

\subsection*{Funding}
This work was supported by MEXT KAKENHI Grant Numbers JP24680031, JP16H05879 and JP25240044 to MH, in part.

\bibliographystyle{natbib}

\bibliography{main}

\clearpage
\appendix
\onecolumn

\renewcommand{\thefigure}{S\arabic{figure}}
\renewcommand{\thesection}{S\arabic{section}}
\renewcommand{\thetable}{S\arabic{table}}
\renewcommand{\theequation}{S\arabic{equation}}
\setcounter{figure}{0}
\setcounter{table}{0}
\setcounter{equation}{0}
\setcounter{section}{0}

\begin{center}
{\LARGE Supplementary Information: \titleStr}
\vspace{3mm}

{\Large Takeda Taikai and Michiaki Hamada}
\end{center}

\section{Alignment accuracy evaluation for insertion states}\label{sec:sup_alignment}

In addition to measuring accuracy according to aligned positions (explained in Section \ref{sec:exp_align_accuracy}), we also evaluated insertion accuracy. We define this measure as precision/recall/f1-score of {\textit{inserted}}. Using the same example as used in Section \ref{sec:exp_align_accuracy}, when the true alignment is $\bfrac{x_1 x_2 x_3 x_4}{- \ y_1 y_2 y_3}$ and the inferred alignment is $\bfrac{x_1 x_2 x_3 x_4}{y_1 \ - y_2 y_3 }$, true insertions and inferred insertions are $\{(x_1, -)\}$ and $\{(x_2, -)\}$. In this case, all of precision/recall/f1 are 0. As shown in Table \ref{tab:f1_ins}, this measure is stricter than traditional measures (Table \ref{tab:f1_pos}) because predicting matched positions is easier than predicting insertions.

\begin{table}[ht]
\caption{Insertion f1 scores \label{tab:f1_ins}} {
\begin{tabular}{@{}lcccccccc@{}} 
\toprule 
              & \multicolumn{8}{c}{Trained models} \\
Simulated model & \texttt{small} & \texttt{med} & \texttt{large} & \texttt{imb} & \texttt{imb\_large} &  \texttt{huge} & \texttt{imb\_huge} & \texttt{fab}\\ 
\midrule 
\texttt{small} & \textit{ 0.5725 } & \textbf{ 0.5764 } & 0.5676 & 0.5689 & 0.5715 & 0.5653 & 0.5677 & 0.5716\\ 
\texttt{med} & 0.7414 & \textit{ 0.7401 } & 0.7375 & \textbf{ 0.7446 } & 0.7371 & 0.7337 & 0.7355 & 0.7404\\ 
\texttt{large} & 0.6365 & 0.6631 & \textit{ 0.6685 } & 0.6506 & 0.6653 & 0.6640 & 0.6695 & \textbf{ 0.6701 }\\ 
\texttt{imb} & 0.6456 & 0.6775 & 0.6733 & \textit{ 0.6767 } & 0.6710 & 0.6731 & 0.6733 & \textbf{ 0.6778 }\\ 
\texttt{imb\_large} & 0.6105 & 0.6422 & \textbf{ 0.6428 } & 0.6330 & \textit{ 0.6427 } & 0.6390 & 0.6393 & 0.6412\\ 
\texttt{huge} & 0.8623 & 0.8736 & 0.8792 & 0.8684 & 0.8759 & \textit{ 0.8825 } & 0.8792 & \textbf{ 0.8821 }\\ 
\texttt{imb\_huge} & 0.9000 & 0.9025 & \textbf{ 0.9111 } & 0.9027 & 0.9104 & 0.9089 & \textit{ 0.9094 } & 0.9091\\ 
\midrule
average & 0.7380 & 0.7494 & 0.7502 & 0.7462 & 0.7500 & 0.7480 & 0.7495 & \textbf{ 0.7514 }\\ 
\botrule 
\end{tabular}
}
{}
\end{table}

\section{Perplexity of trained models}\label{sec:perplexity}

Perplexity is a measure of how accurately a probabilistic model predicts a sample (in this case a sample properly corresponding to an alignment) and is defined as 
\begin{align}
    \text{Perplexity}
    &= \exp \left\{-\frac{1}{N} \ln \sum_{\vec{Z}} p(\vec{X}_{\text{held-out}}, \vec{Z} | \vec{\Pi}, \mathcal{M})\right\} \\
    &= \exp \left\{-\frac{1}{N} \ln p(\vec{X}_{\text{held-out}} | \vec{\Pi}, \mathcal{M})\right\}
\end{align}
where $\vec{X}_{\text{held-out}}$ represents the reserved sequences (i.e. sequences not used for training) and $N$ is the number of sequences (500 sequences in this experiment). 

Because perplexity can be very large, we report log perplexity instead (Supplementary Table \ref{tab:perplexity}). The proposed method outperformed smaller trained models when simulated models were large. For example, when the simulated model was \texttt{huge}, our proposed method achieved a log perplexity of 177.0064 but the \texttt{med} model's log perplexity is only 179.7014. 
Because \texttt{small} or \texttt{med} models are usually considered for alignment tasks, our proposed model has advantages for alignment modelling over existing methods. 
It is unexpected but interesting that larger models (such as \texttt{huge} and \texttt{large}) achieve accuracy that is comparable to that of the proposed method, which indicates that the larger models do not result in overfitting even when the data are generated by simpler models.

\begin{table*}[ht]
\caption{Log perplexity for reserved data (smaller values are better). For each model used for data generation (i.e. the simulated model), we trained the model with fixed model sizes including the true model size and the proposed model (\texttt{fab}) before the perplexity evaluation. Italic and bold values indicates perplexity scores of models trained with the true model size and the best perplexity score except for that inferred using the true model size, respectively. \label{tab:perplexity}} 
\centerline{
\begin{tabular}{@{}ccccccccc@{}} 
\toprule 
trained model & small & med & large & imbalanced & imbalanced\_large & huge & imbalanced\_huge & fab\\ 
\midrule 
small & \textit{ 217.9860 } & 217.9880 & 217.9926 & \textbf{ 217.9841 } & 217.9917 & 217.9929 & 217.9965 & 217.9860\\ 
med & 204.7439 & \textit{ 204.6744 } & \textbf{ 204.6718 } & 204.7240 & 204.6725 & 204.6769 & 204.6820 & 204.6755\\ 
large4 & 214.7708 & 213.2542 & \textit{ 212.8900 } & 214.0971 & 213.0589 & 212.8999 & 212.8955 & \textbf{ 212.8900 }\\ 
imbalanced & 215.5882 & 214.2698 & 214.2883 & \textit{ 214.2692 } & 214.2832 & 214.2871 & 214.2905 & \textbf{ 214.2694 }\\ 
imbalanced\_large & 216.9349 & 216.0192 & 215.8606 & 216.1357 & \textit{ 215.8605 } & 215.8620 & 215.8781 & \textbf{ 215.8597 }\\ 
huge & 182.5948 & 179.7014 & 177.1100 & 180.9010 & 178.1934 & \textit{ 177.0079 } & 177.0951 & \textbf{ 177.0064 }\\ 
imbalanced\_huge & 175.9397 & 172.7472 & 169.9543 & 174.1477 & 171.1988 & \textbf{ 169.8718 } & \textit{ 169.8713 } & 169.8721\\ 
ave & 200.2835 & 198.8588 & 198.0205 & 199.2063 & 198.2580 & 197.9942 & 198.0045 & \textbf{ 197.9884 }\\ 
\botrule 
\end{tabular}
}
{}
\end{table*}


\section{Parameters for artificial models}\label{sec:params_art}

\begin{figure}[ht]
  \centering
\begin{tabular}{cc}
(a) Initial/transition probability &  (b) Emission probability \\
  \includegraphics[width=.5\linewidth]{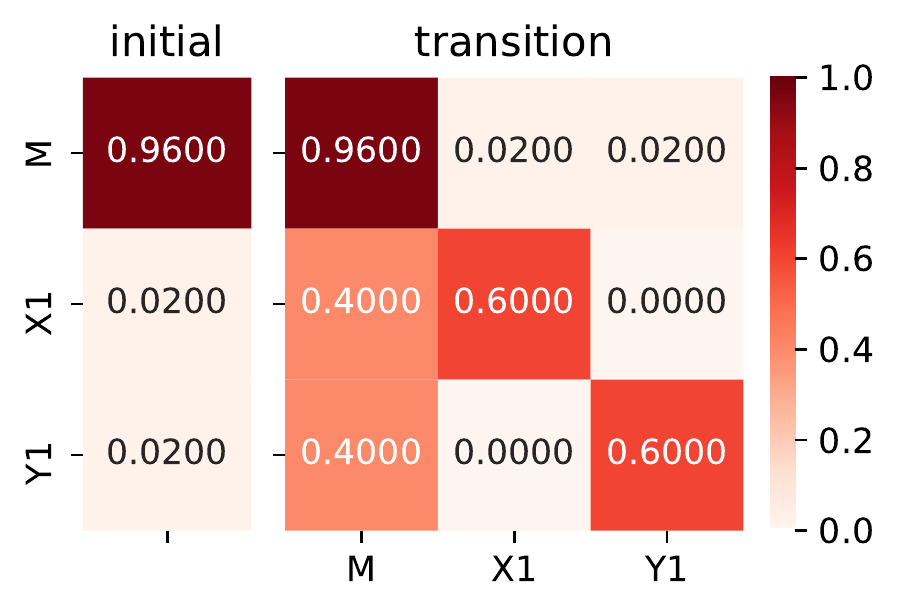} & 
  \includegraphics[width=.5\linewidth]{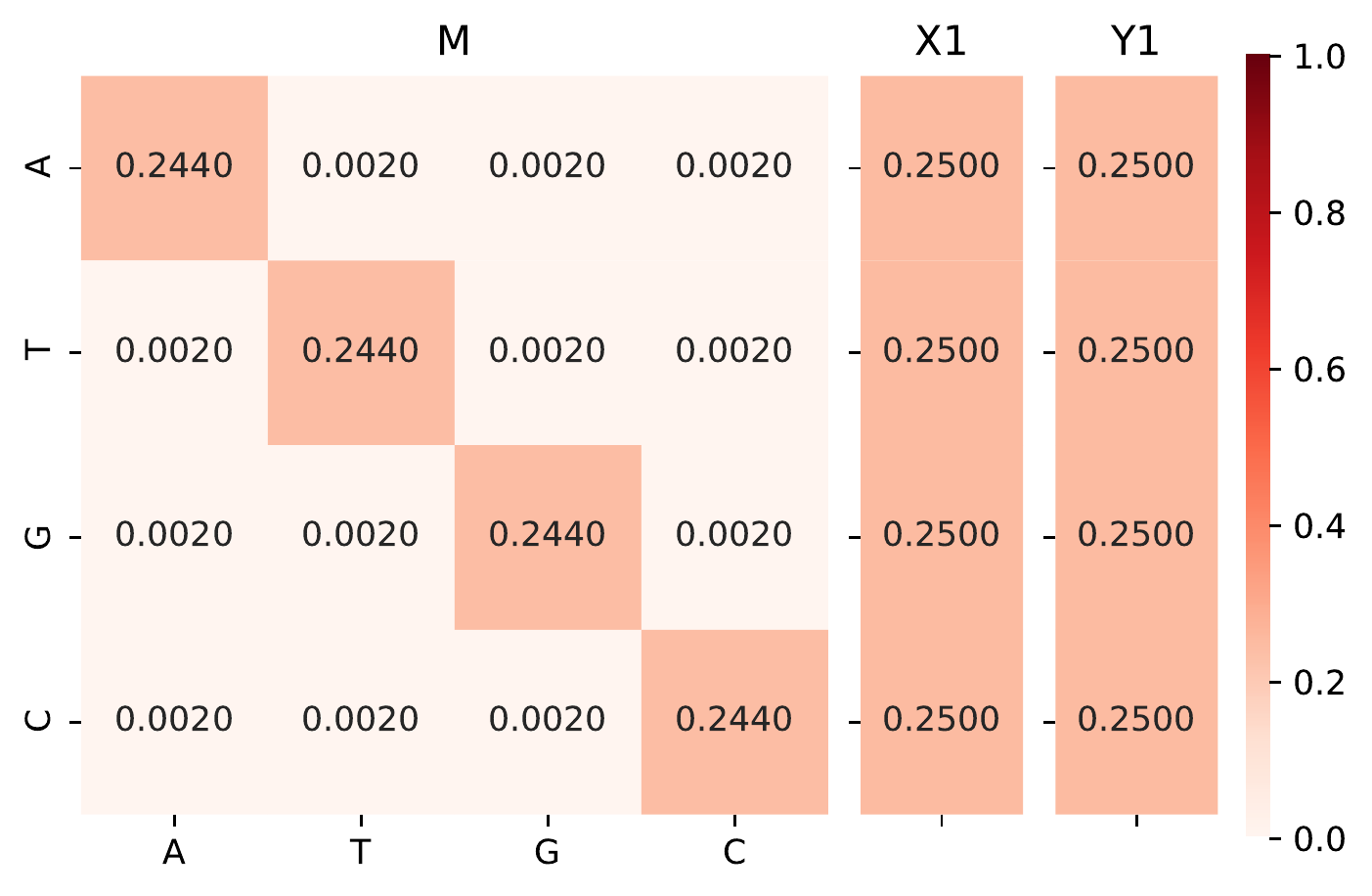}
  \end{tabular}
  \caption{Parameters of \texttt{small} model}  
  \label{fig:params_small}
\end{figure}

\begin{figure}[ht]
  \centering
\begin{tabular}{cc}
(a) Initial/transition probability &  (b) Emission probability \\
  \includegraphics[width=.5\linewidth]{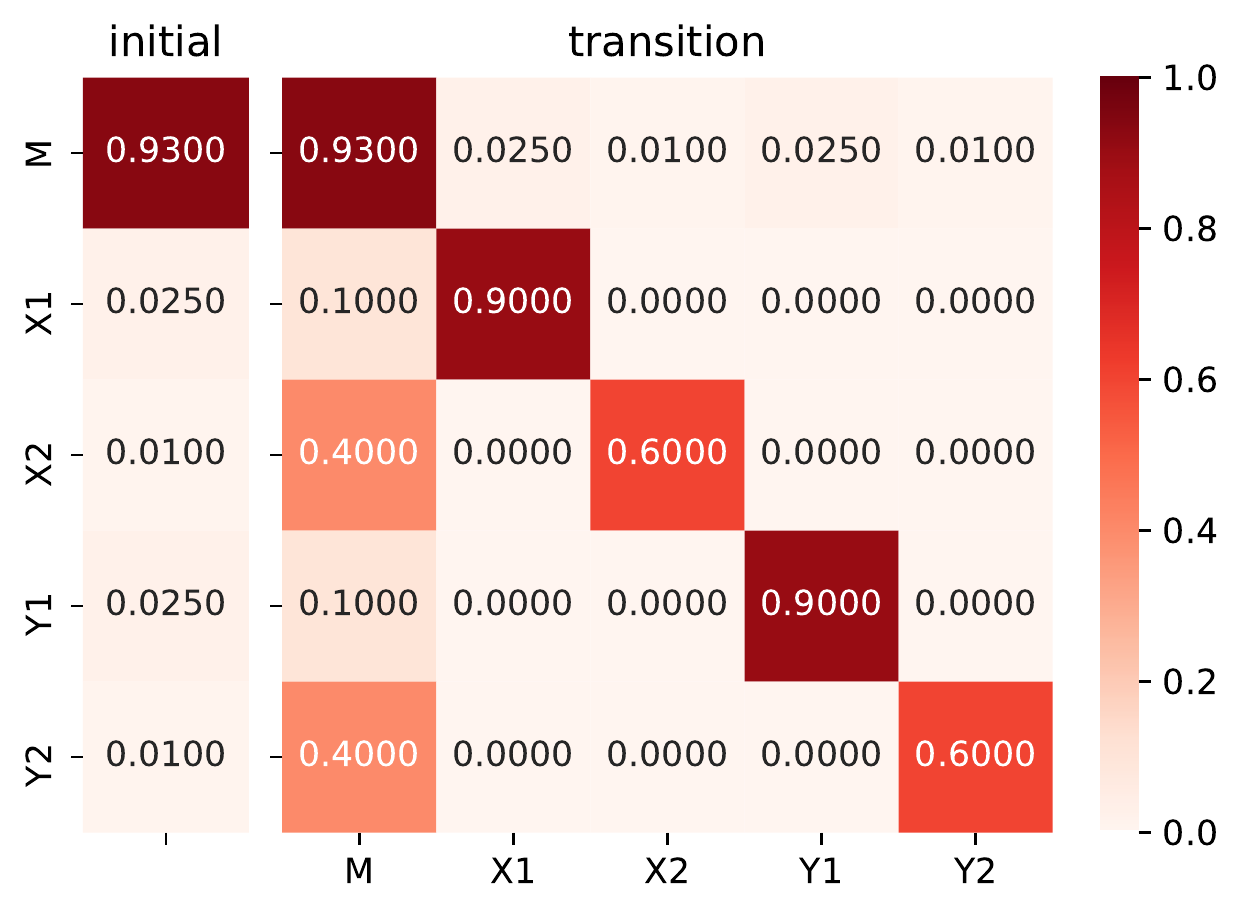} & 
  \includegraphics[width=.5\linewidth]{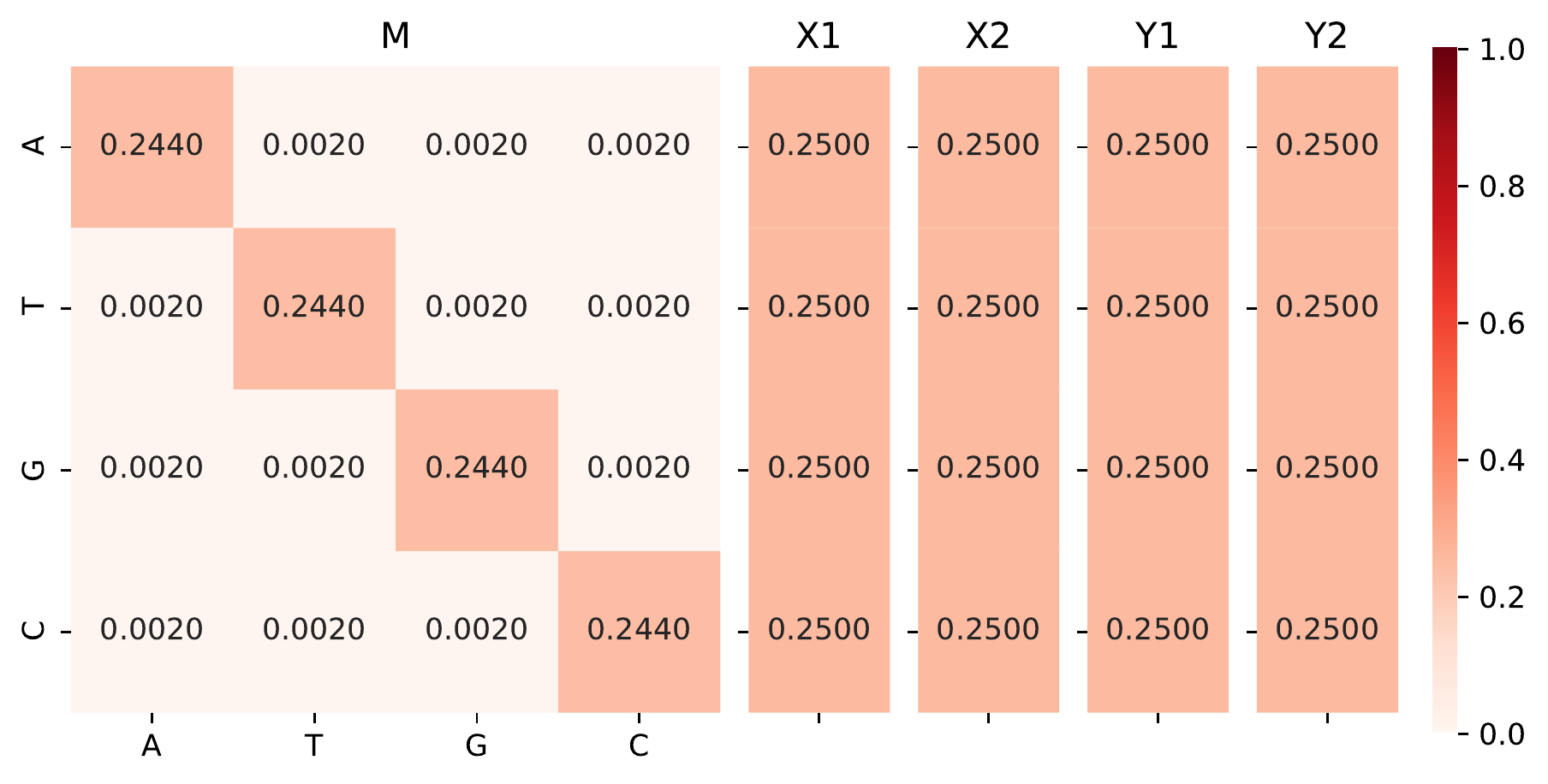}
  \end{tabular}
  \caption{Parameters of \texttt{med} model}  
  \label{fig:params_med}
\end{figure}

\begin{figure}[ht]
  \centering
\begin{tabular}{cc}
(a) Initial/transition probability &  (b) Emission probability \\
  \includegraphics[width=.5\linewidth]{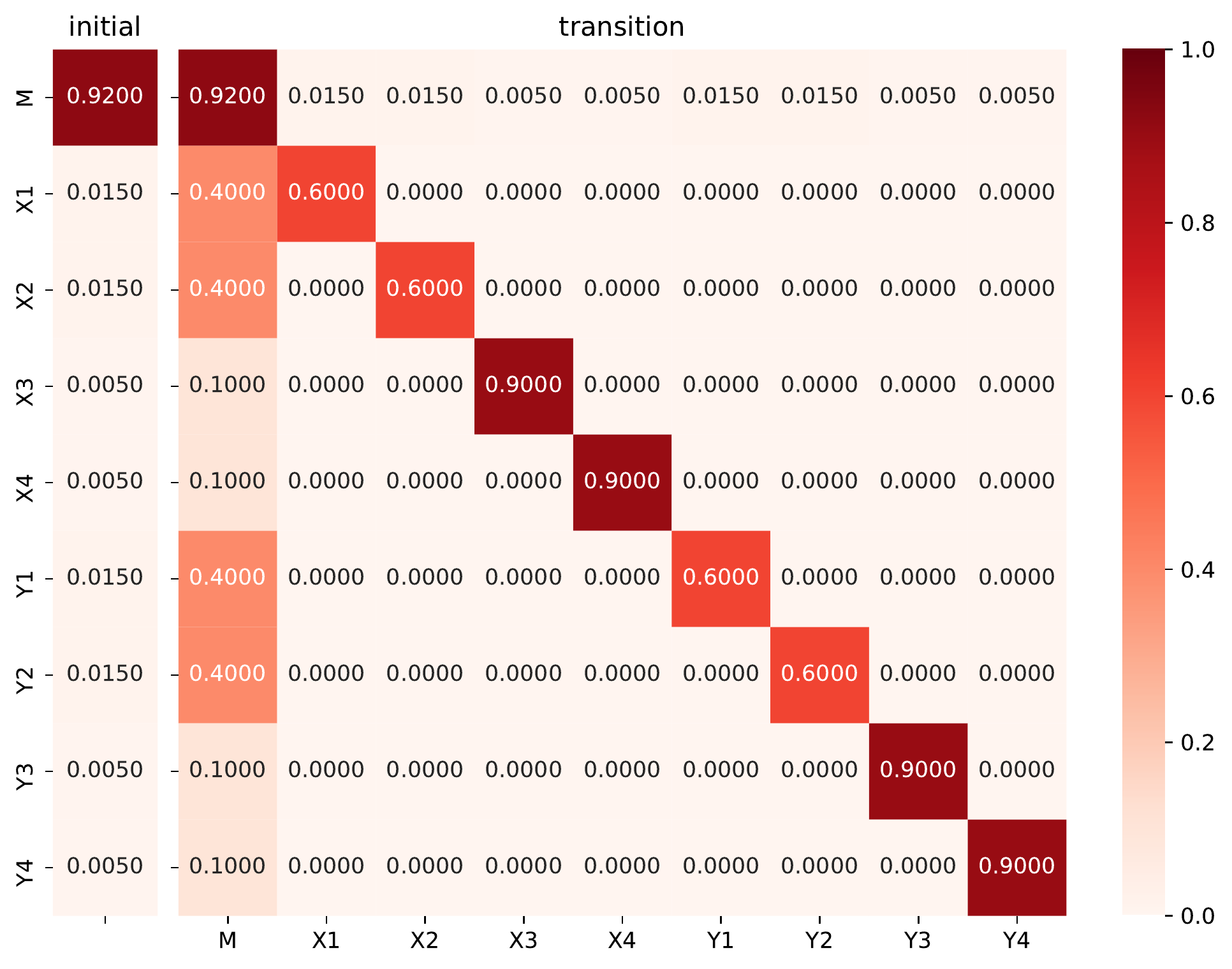} & 
  \includegraphics[width=.5\linewidth]{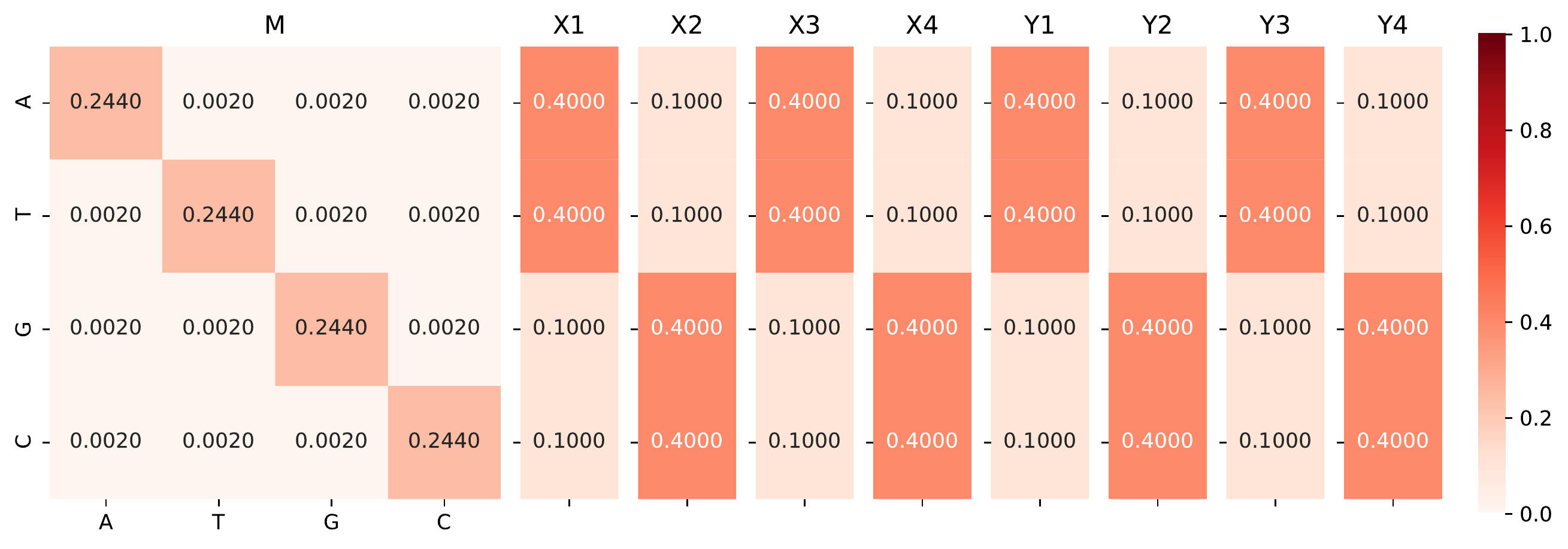}
  \end{tabular}
  \caption{Parameters of \texttt{large} model}  
  \label{fig:params_large}
\end{figure}

\begin{figure}[ht]
  \centering
\begin{tabular}{cc}
(a) Initial/transition probability &  (b) Emission probability \\
  \includegraphics[width=.5\linewidth]{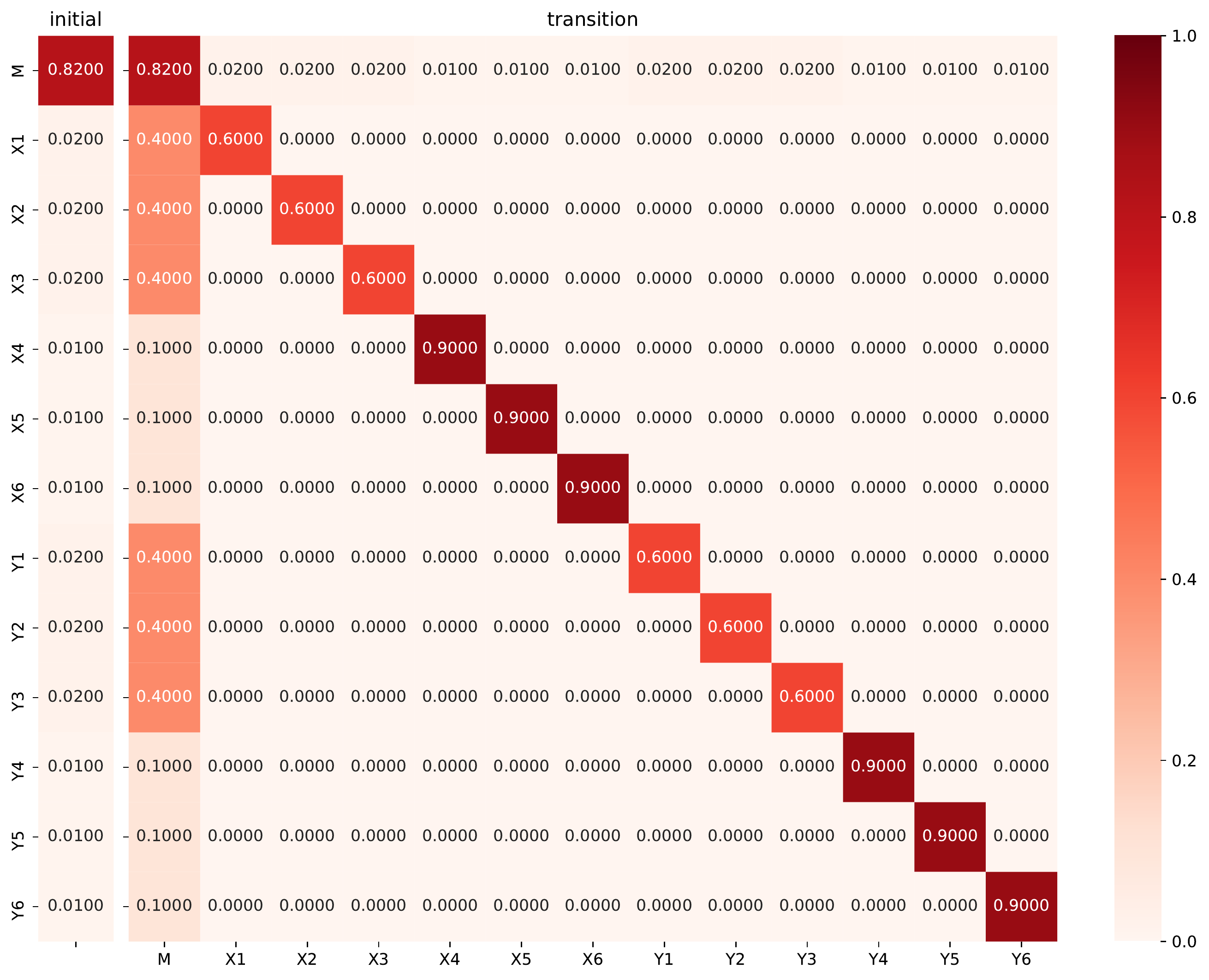} & 
  \includegraphics[width=.5\linewidth]{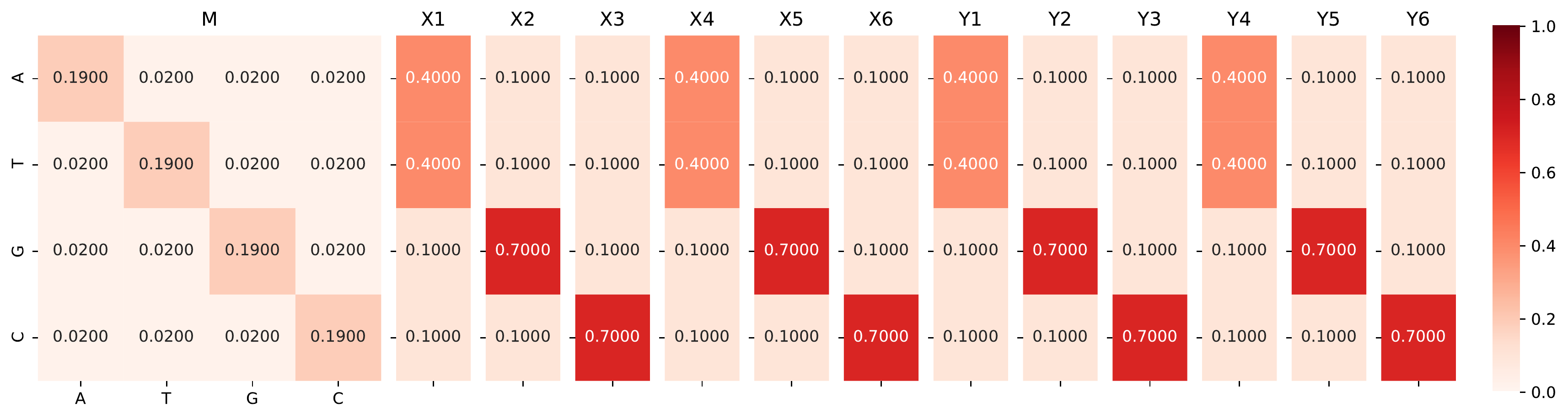}
  \end{tabular}
  \caption{Parameters of \texttt{huge} model}  
  \label{fig:params_huge}
\end{figure}

\begin{figure}[ht]
  \centering
\begin{tabular}{cc}
(a) Initial/transition probability &  (b) Emission probability \\
  \includegraphics[width=.5\linewidth]{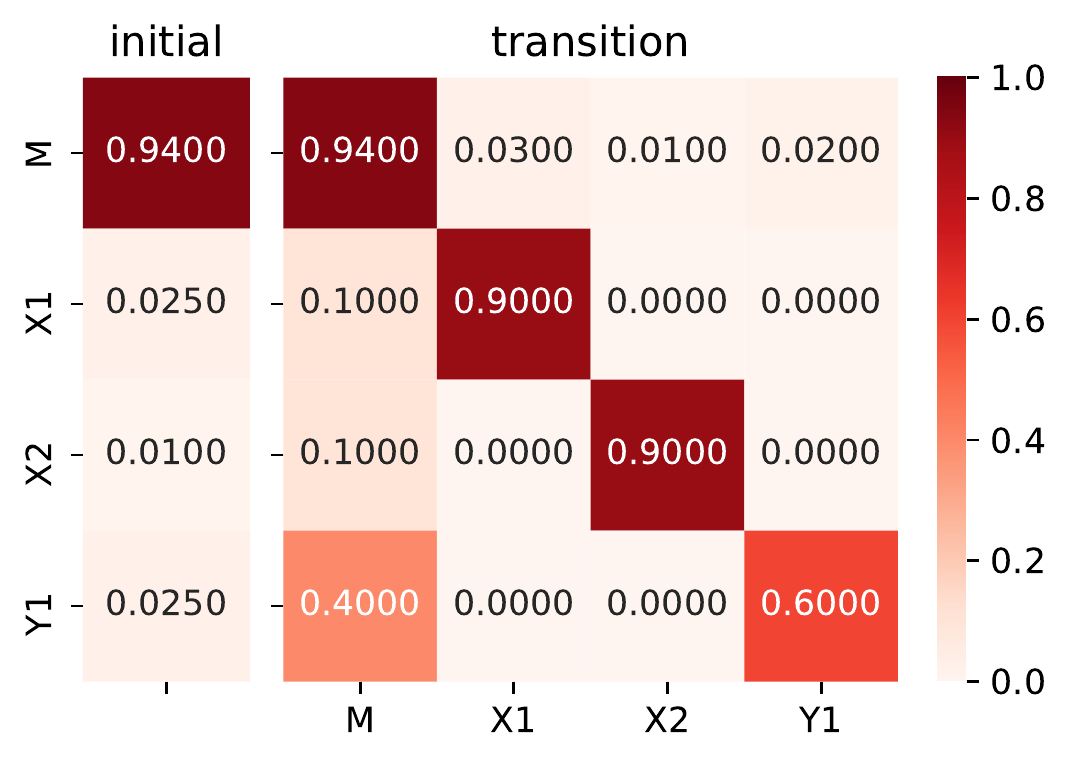} & 
  \includegraphics[width=.5\linewidth]{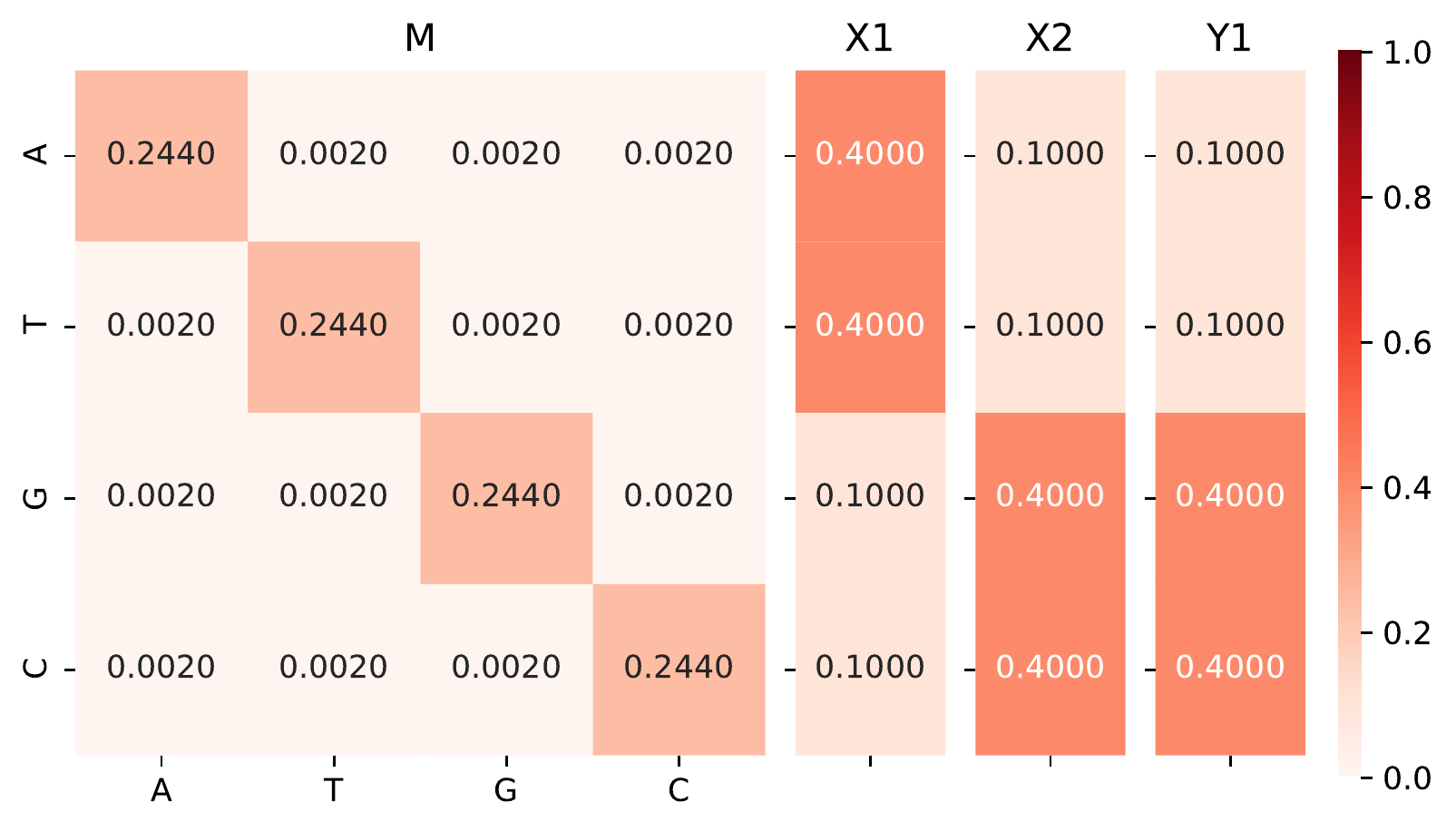}
  \end{tabular}
  \caption{Parameters of \texttt{imb} model}  
  \label{fig:params_imb}
\end{figure}

\begin{figure}[ht]
  \centering
\begin{tabular}{cc}
(a) Initial/transition probability &  (b) Emission probability \\
  \includegraphics[width=.5\linewidth]{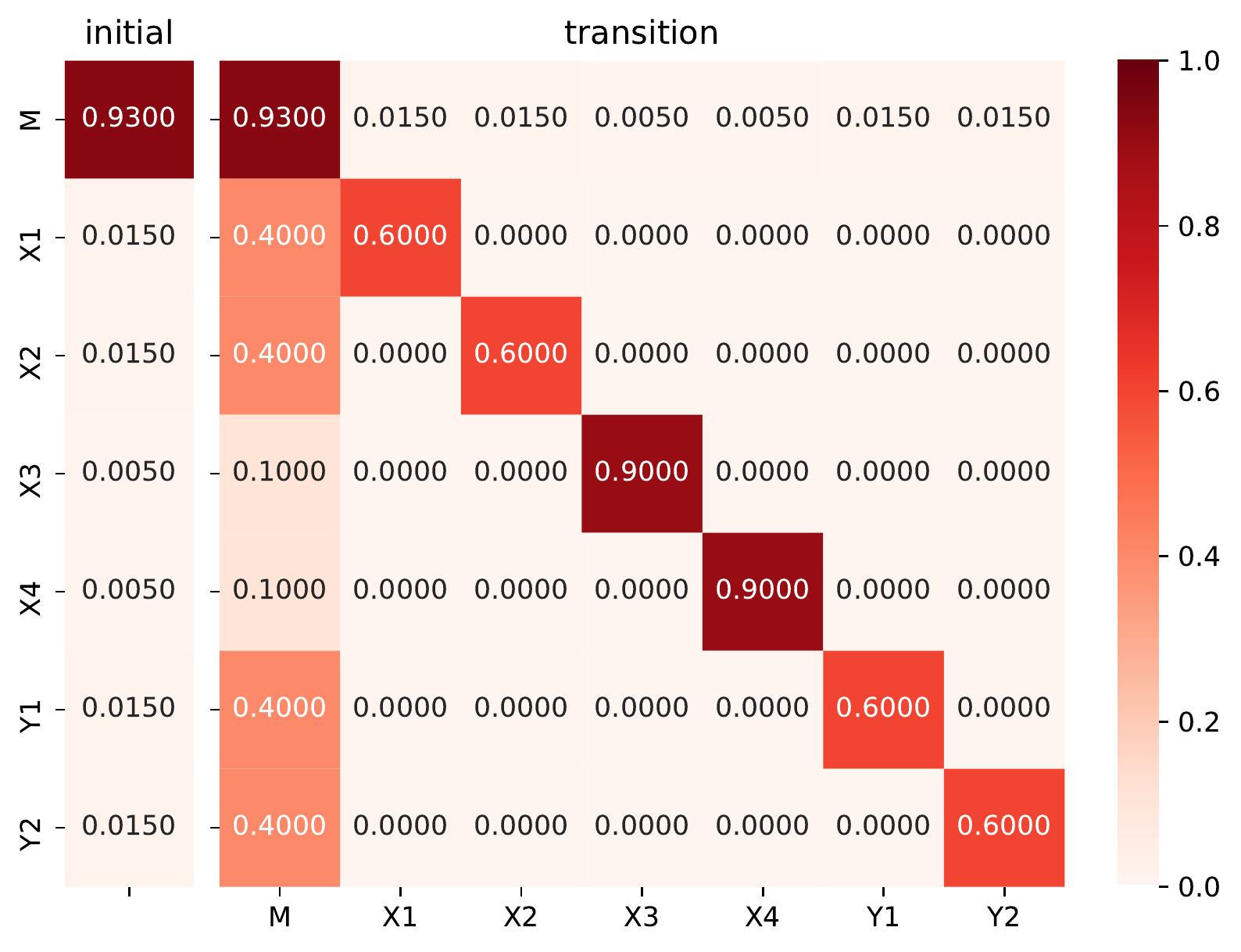} & 
  \includegraphics[width=.5\linewidth]{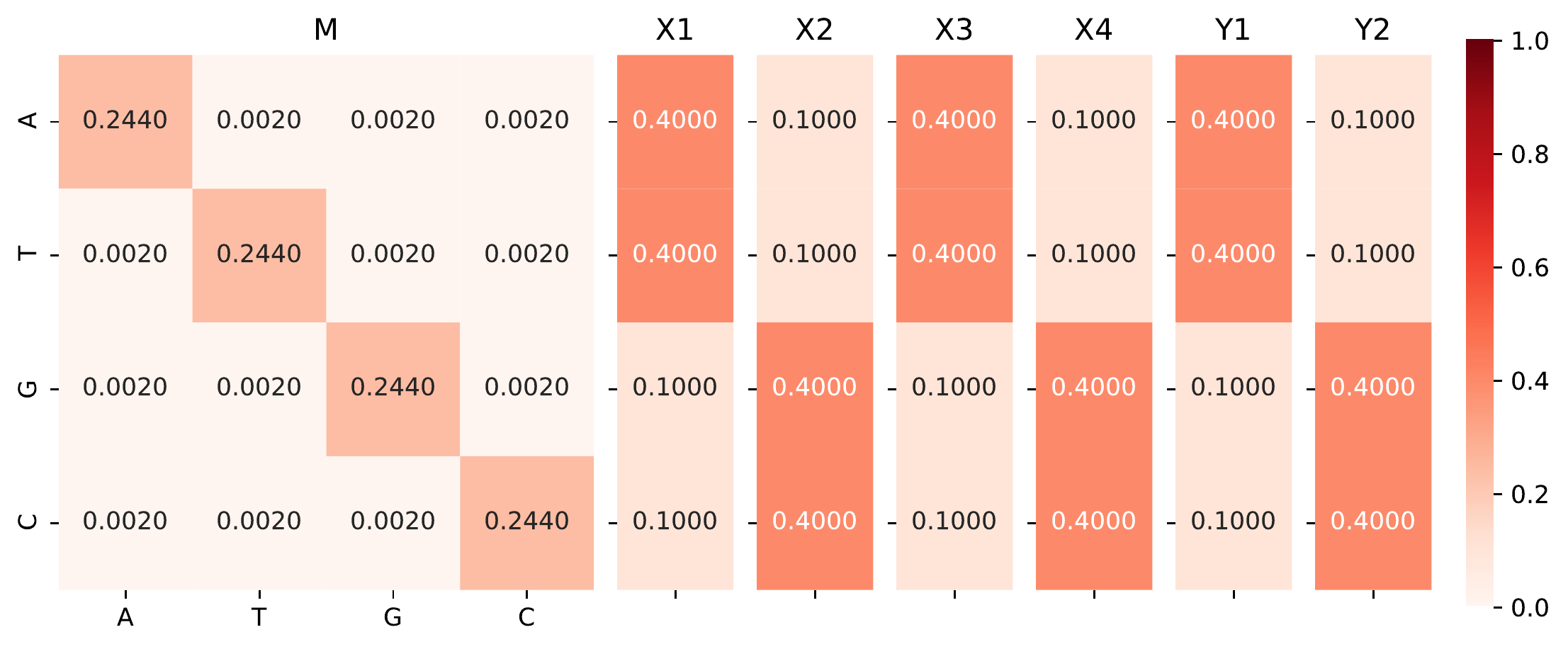}
  \end{tabular}
  \caption{Parameters of \texttt{imb\_large} model}  
  \label{fig:params_imb_large}
\end{figure}

\begin{figure}[ht]
  \centering
\begin{tabular}{cc}
(a) Initial/transition probability &  (b) Emission probability \\
  \includegraphics[width=.5\linewidth]{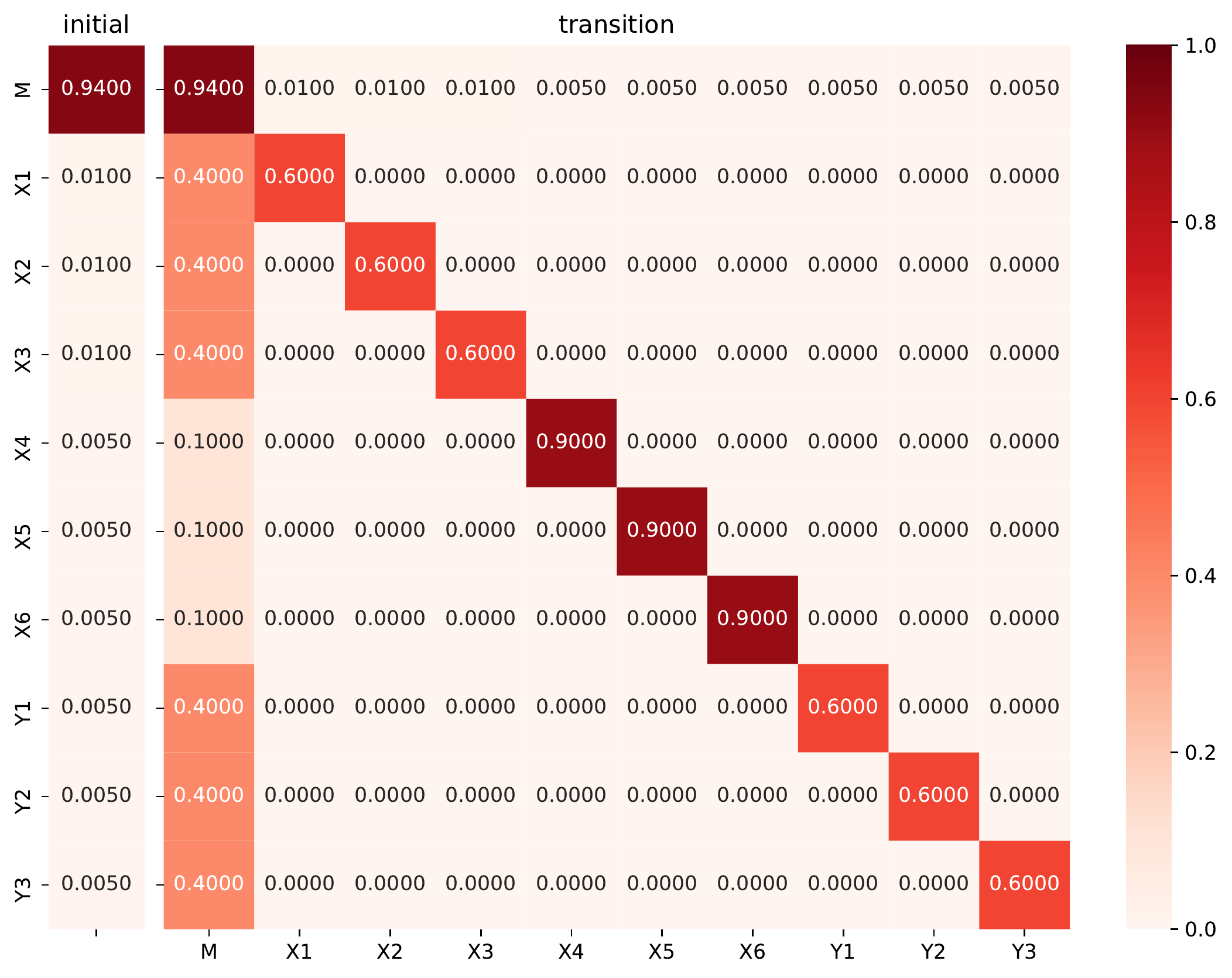} & 
  \includegraphics[width=.5\linewidth]{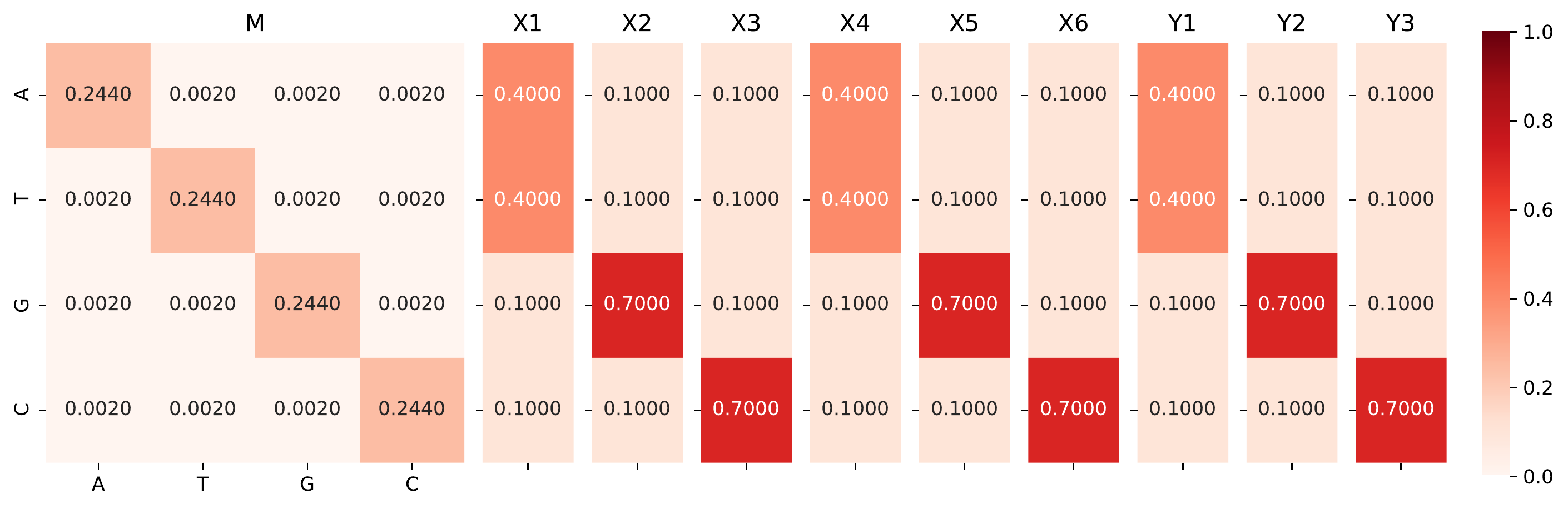}
  \end{tabular}
  \caption{Parameters of \texttt{imb\_huge} model}  
  \label{fig:params_imb_huge}
\end{figure}

Here, we show initial/transition/emission probabilities of each artificial model in Figure \ref{fig:params_small} (\texttt{small}), \ref{fig:params_med} (\texttt{med}),  \ref{fig:params_large} (\texttt{large}),  \ref{fig:params_huge} (\texttt{huge}),  \ref{fig:params_imb} (\texttt{imb}),  \ref{fig:params_imb_large} (\texttt{imb\_large}) and  \ref{fig:params_imb_huge} (\texttt{imb\_huge}).

\section{Parameter visualisation of the model inferred from the LAST dataset}\label{sec:last_params}
\begin{figure}[ht]
  \centering
\begin{tabular}{c}
(a) Initial/transition probability\\  
  \includegraphics[width=0.7\linewidth]{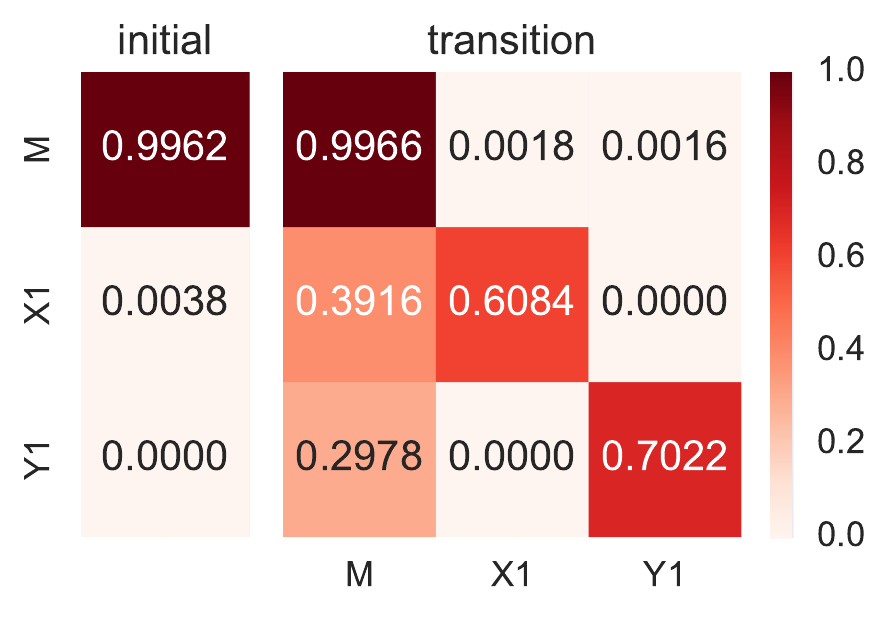}
\\
 (b) Emission probability \\  
  \includegraphics[width=0.8\linewidth]{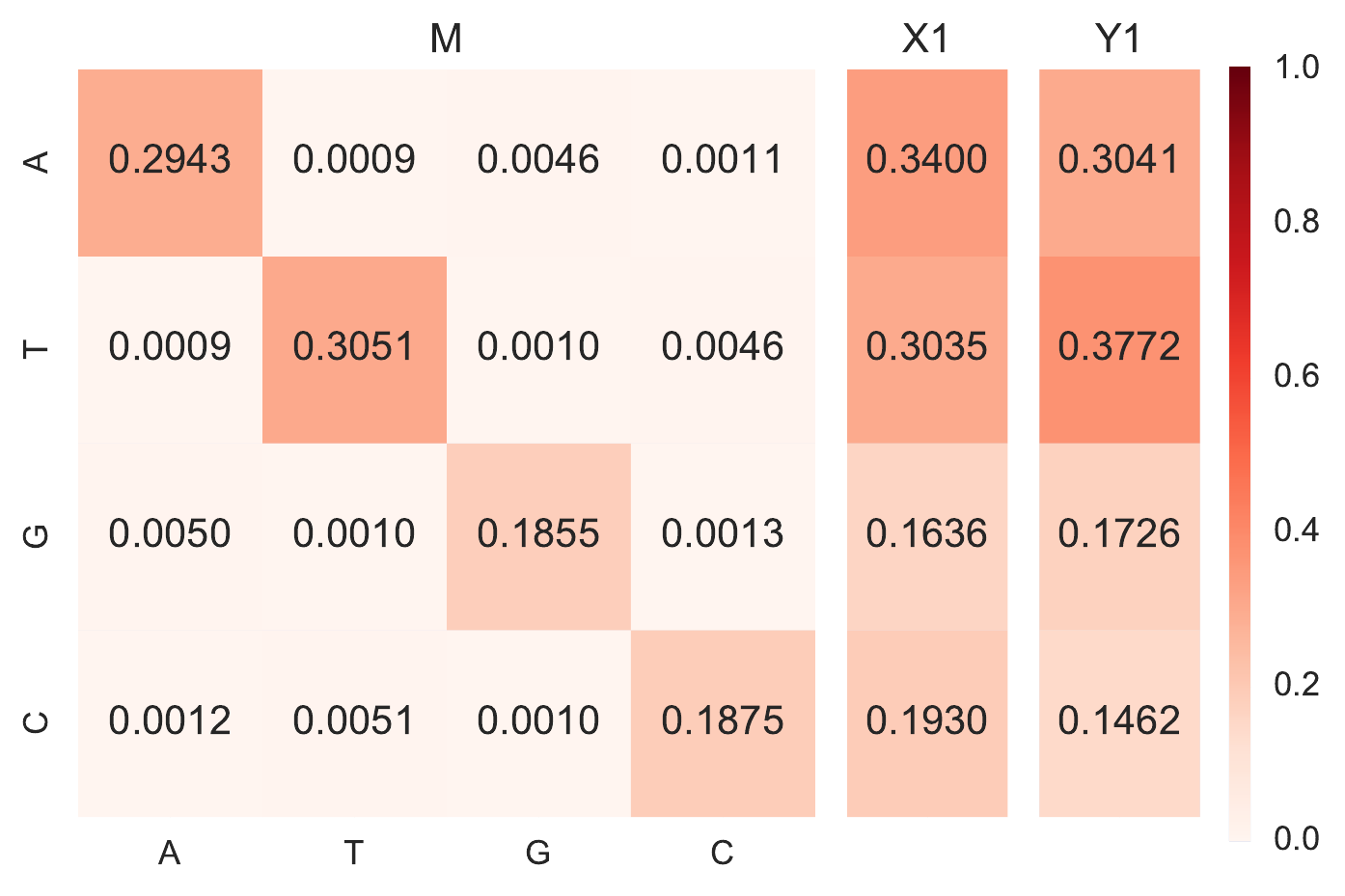}
  \end{tabular}
    
    \caption{Trained PHMM and its parameters for alignments between human and orangutan sequences. As observed in the human--chimpanzee alignment (Fig. \ref{fig:params_chimp}), the resulting model is the simplest one, $(K_M, K_X, K_Y) = (1, 1, 1)$.}
  \label{fig:params_orangutan}
\end{figure}

\begin{figure}[ht]
  \centering
\begin{tabular}{c}
(a) Initial/transition probability\\
  \includegraphics[width=\linewidth]{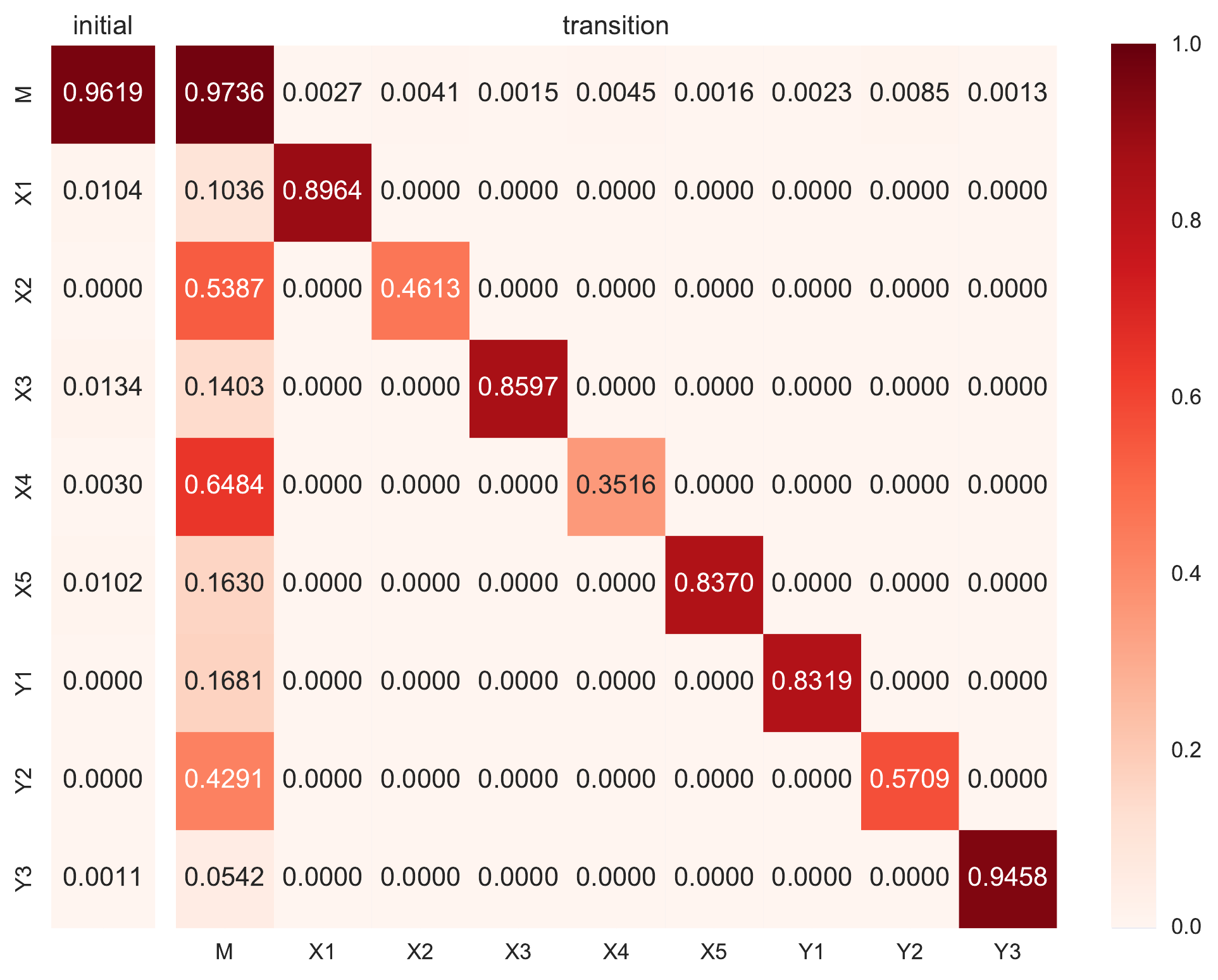}
\\
 (b) Emission probability \\
  \includegraphics[width=\linewidth]{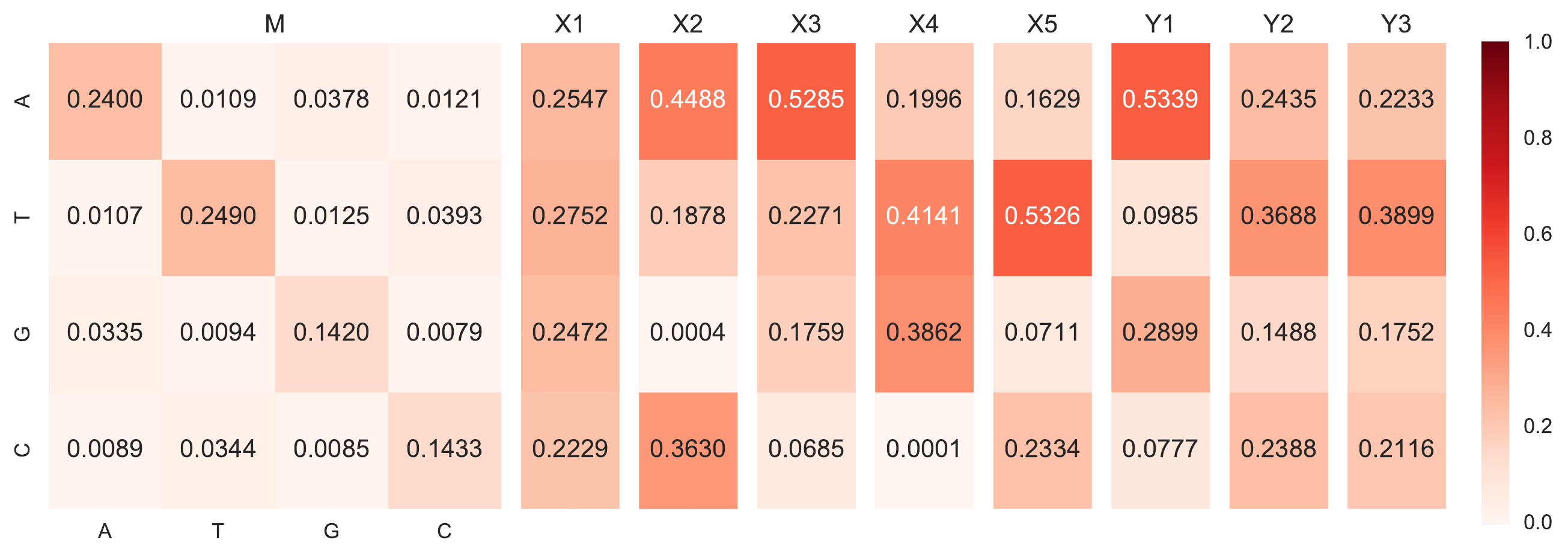}

  \end{tabular}
  \caption{Trained PHMM and its parameters for the alignment between human and dog sequences.
  In (a) $X1$--$X5$ and $Y1$--$Y3$ show five $X$-insertion states and three $Y$-insertion states, respectively, where each value is an estimated initial/transition probability; for example, the value of $0.0041$ in cell $(M, X2)$ is equal to the transition probability from $M$ to $X2$.}  
  \label{fig:params_dog}
\end{figure}

\end{document}